\documentclass[aps,pra, reprint, nofootinbib]{revtex4-1}

\usepackage{graphicx} 
\usepackage{ae,aecompl} 
\usepackage{bm} 
\usepackage{amsmath} 
\usepackage{amssymb} 
\usepackage{textcomp} 
\usepackage{xcolor} 
\usepackage{soul}

\newcommand{\vctr}[1]{\ensuremath{\mathbf{ #1 }}}

\newcommand{\pbp}[2]{\ensuremath{\frac{
\partial #1}{
\partial #2}}} 
 
\newcommand{\e}{\mathrm{e}}
\newcommand{\ket}[1]{\ensuremath{\left| #1 \right\rangle}} 
\newcommand{\bra}[1]{\ensuremath{\left\langle #1 \right|}} 
\newcommand{\bk}[2]{\ensuremath{\left\langle #1 | #2 \right\rangle}} 
\newcommand{\mean}[1]{\ensuremath{\langle #1 \rangle}} 
\newcommand{\abs}[1]{\ensuremath{\left| #1 \right|}} 
\newcommand{\proj}[2]{\ensuremath{\ket{#1} \bra{#2}}}

\newcommand{\matel}[3]{\ensuremath{\bra{#1} \op{#2} \ket{#3}}} 
 
\newcommand{\op}[1]{\ensuremath{\hat{\textsf{\ensuremath{#1}}}}} 
\newcommand{\opad}[1]{\ensuremath{\op{#1}^{\dagger}}}

\newcommand{\vphi}{\ensuremath{\varphi}} 
\newcommand{\Jz}{\ensuremath{\op{J_{z}\,}}} 
\newcommand{\Jx}{\ensuremath{\op{J_{x}\,}}} 
\newcommand{\Jy}{\ensuremath{\op{J_{y}\,}}} 
 
\newcommand{\D}[1]{\ensuremath{\Delta #1^{2} \,}} 
\newcommand{\xis}{\ensuremath{\xi_{S}^{2}\,}} 
\newcommand{\xin}{\ensuremath{\xi_{N}^{2}\,}}

\begin{document}

\title[Spin squeezing, entanglement and quantum metrology with Bose-Einstein condensates.] {Spin squeezing, entanglement and quantum metrology with Bose-Einstein condensates.}

\author{Christian Gross} 
\affiliation{Kirchhoff-Institut f\"ur Physik, Universit\"at Heidelberg, Im Neuenheimer Feld 227, 69120 Heidelberg, Germany} 
\altaffiliation{Present address: Max-Planck-Institut f\"ur Quantenoptik, Hans-Kopfermann-Strasse 1, 85748 Garching, Germany}
\email{christian.gross@mpg.mpq.de}

\begin{abstract}
Squeezed states, a special kind of entangled states, are known as a useful resource for quantum metrology. 
In interferometric sensors they allow to overcome the ``classical'' projection noise limit stemming from the independent nature of the individual photons or atoms within the interferometer.
Motivated by the potential impact on metrology as wells as by fundamental questions in the context of entanglement, a lot of theoretical and experimental effort has been made to study squeezed states. The first squeezed states useful for quantum enhanced metrology have been proposed and generated in quantum optics, where the squeezed variables are the coherences of the light field. 
In this tutorial we focus on spin squeezing in atomic systems. We give an introduction to its concepts and discuss its generation in Bose-Einstein condensates. We discuss in detail the experimental requirements necessary for the generation and direct detection of coherent spin squeezing. Two exemplary experiments demonstrating adiabatically prepared spin squeezing based on motional degrees of freedom and diabatically realized spin squeezing based on internal hyperfine degrees of freedom are discussed. 
\end{abstract}

\maketitle

\section{Introduction.} \label{sec.intro}

Interferometers, which are among the most precise measurement devices known today, are based on the interference of two coherent atomic or photonic modes. 
The quantity to be measured is mapped to the relative phase between the two modes, explicitly using the wave-like nature of the resources (atoms or photons). However, in the actual detection process their particle-like character becomes important; the population of the two modes can be at best be counted one by one. Traditional interferometers use uncorrelated resources in two coherent modes which results in a Poissonian distribution of the population difference at the interferometer output. The standard deviation of this distribution $\sqrt{N}$ is the so called projection noise or shot noise limit.

The best theoretically possible phase precision for uncorrelated resources and linear interferometers is given by the classical Cramer-Rao bound of parameter estimation~\cite{Cramer1946, Braunstein1994}. In the context of (classical) interferometric phase estimation this is known as the standard quantum limit. It can be reached with coherent states, the quantum states that are naturally generated by coherent splitting of a quantum mode, as it is routinely done by beam splitters in optics (with one input port unused). Indeed, modern interferometers measuring for example time~\cite{Santarelli1999}, position~\cite{Goda2008, Arcizet2006} or magnetic fields~\cite{Wasilewski2010} are limited by this quantum noise leading to a maximal interferometric phase precision of $\Delta\varphi = 1/\sqrt{N}$. 
 
Several strategies have been proposed to overcome this standard quantum limit for metrology by using quantum correlated, that is entangled, resources within the interferometer~\cite{Giovannetti2006a}. 
One of these strategies is to use squeezed quantum states~\cite{Walls1983} which have been successfully generated in quantum optics~\cite{Slusher1985a}.
The analogue for atom interferometry are spin squeezed states~\cite{Kitagawa1993, Wineland1994}. 
Spin squeezed states can be obtained from coherent spin states by a continuous transformation which gradually redistributes the quantum noise among two orthogonal spin directions. This property makes them promising candidates for their experimental realization since the degree of squeezing can be balanced with the increased sensitivity to decoherence. Interferometric precision is increased to $\Delta\varphi = \xi_S/\sqrt{N}$ for spin squeezed states with a squeezing parameter $\xi_S$.

Many-body quantum mechanics sets an ultimate limit to the phase precision for general interferometric measurements. This is the so-called Heisenberg limit, for which an $1/N$ scaling of the interferometric precision can be achieved~\cite{Giovannetti2006a}. The Heisenberg limit can be reached with spin squeezed states in the limit of infinite squeezing where spin squeezed states become identical to twin Fock states. 
The potential gain in phase precision is enormous: In an atomic clock, for example, one measures an energy difference $\hbar \omega$ for a certain time $t$ ($2\pi\hbar$ is Planck's constant). The ``classical'' uncertainty scales as $\D{\omega} = 1/t N$, while Heisenberg limited metrology would allow for an error of $\D{\omega} = 1/t N^{2}$. Assuming a fictitious Heisenberg limited measurement with $10^{6}$ atoms lasting one second, a ``classical'' projection noise limited apparatus would need $11$ days to obtain the same level of precision. \\
In this tutorial we attempt to provide a basic introduction into the concept of spin squeezing. We focus on the theoretical concept of spin squeezing, its connection to entanglement and applications in quantum metrology. Experimental techniques to generate and detect spin squeezing in Bose-Einstein condensates (BEC) are explained in detail. Finally we shortly discuss two exemplary experiments in which spin squeezing has been observed in BECs.

\section{The concept of spin-squeezing.} Spin squeezing is a quantum strategy introduced in 1993 by Kitagawa and Ueda~\cite{Kitagawa1993} which aims to redistribute the fluctuations of two orthogonal spin directions among each other. In 1994 it was theoretically shown that spin squeezed states are useful quantum resources to enhance the precision of atom interferometers~\cite{Wineland1994} and in 2001 the connection between spin squeezing and entanglement was pointed out~\cite{Sørensen2001}. In the field of quantum optics squeezed states have been proposed to overcome the standard quantum limit in optical interferometry already in 1981~\cite{Caves1981, Walls1983}. These states are similar to spin squeezed states however differences arise from the different variables, quadratures versus collective spin directions~\cite{Wang2003a}.
The connection of spins to interferometry becomes obvious when realizing that an interferometer is essentially a two mode system with modes $\ket{a}$ and $\ket{b}$. Each particle in the interferometer can be regarded as an elementary spin with spin $\sigma =1/2$, with two possible longitudinal  eigenstates $\sigma_z = \pm 1/2$ corresponding to the two interferometer modes. 
The relative phase measured by the interferometer corresponds to the spin component in the plane spanned by the transversal spin directions $\sigma_x$ and $\sigma_y$. 
The generalization of this picture to $N$ particles, which results in large collective spin vectors $\op{J}$, is outlined below. 
An extensive theoretical review of two-mode systems also detailing the pseudo spin description can be found in~\cite{Dalton2012}.

\subsection{Collective spins.}

The mathematical concept of a spin algebra with total spin $J$ is a powerful tool to describe very different physical systems. Any observable within such a spin $J$ system can be expressed by the three spin operators $\Jx$, $\Jy$, $\Jz$ and the identity operator. The collective spin operators $\op{J}_i$ can be defined as the sum over all elementary spin operators (Pauli matrices) $\op{\sigma}_{i}^{(k)}$, where $i=(x, y, z)$: 
\begin{equation}
	\op{J}_i= \sum_{k=1}^{N} \op{\sigma}_{i}^{(k)} 
	\label{eq.spingeneral} 
\end{equation}

The three orthogonal spin components are non-commuting operators. Their commutation relation is $[\op{J}_{i}, \op{J}_{j}]=i \,\epsilon_{ijk} \op{J}_{k}$, where $\epsilon_{ijk}$ is the Levi-Civita symbol. Therefore, any pair of spin operators obeys a Heisenberg uncertainty relation which, for $\D{\Jz}$ and $\D{\Jy}$, is given by 
\begin{equation}
	\D{\Jz}\D{\Jy}\geq\frac{1}{4}{\mean{\Jx}}^{2} 
	\label{eq.spinheisenberg}
\end{equation}
and $\D{\Jz}=\mean{\Jz^{2}}-\mean{\Jz}^{2}$ is the variance in $J_z$ direction. 

A basis of the general problem can be obtained as the tensor product of all $N$ bases of the individual components, each of dimension $(2\sigma+1)$ -- the dimension of the Hilbert space is huge $\mathrm{dim}(\mathcal{H_{N}})=(2\sigma+1)^{N} = 2^{N}$ and grows exponentially with the number of elementary spins. The length of the collective spin $J$ is smaller or equal than half the number of elementary spins. The spin length in interferometry is usually large such that  $\sqrt{J(J+1)} \approx J \leq N/2$ is a good approximation. 

In the symmetric case, that is, when all operations done on the spin ensemble affect each spin in the same way (such that there is exchange symmetry) the problem simplifies considerably. For example, each elementary spin can be prepared in the $\sigma_{z}=-1/2$ state and maximum collective polarization $J_{z}=-N/2$ can be reached. Hence, the total spin length is given by $J = N/2$ and the dimension of the Hilbert space dramatically reduces to $\mathrm{dim}(\mathcal{H_{S}})=(2N\sigma+1) = (N+1)$, linearly dependent on the number of elementary spins. One possible choice of a basis are the symmetric Dicke states~\cite{Arecchi1972} $\ket{J,m}$ with $-N/2<m<N/2$. Each of the Dicke states corresponds to a perfectly defined particle number difference between the two modes $\ket{a}$ and $\ket{b}$ and -- since the total number of particles $N$ is fixed -- the Dicke states correspond to twin Fock states in the modes $\ket{a}$ and $\ket{b}$.

Due to their exchange symmetry the elementary spins can be effectively described as Bosons, the Schwinger Bosons~\cite{Sakurai2004}. Employing the second quantization formalism the creation ($\opad{a}$, $\opad{b}$) and annihilation operators ($\op{a}$, $\op{b}$) of the two modes can be related to the different spin components~\cite{Sakurai2004} 
\begin{eqnarray*}
	\op{J}_{+} &=& \opad{b}\op{a} \nonumber \\
	\op{J}_{-} &=& \opad{a}\op{b} \nonumber \\
	\Jx &=& \frac{1}{2}(\op{J}_{+} + \op{J}_{-}) \nonumber \\
	\Jy &=& \frac{1}{2i}(\op{J}_{+} - \op{J}_{-}) \nonumber \\
	\Jz &=& \frac{1}{2}(\op{n}_b-\op{n}_a) 
	\label{eq.Schwinger} 
\end{eqnarray*}
with $\op{n}_a = \opad{a}\op{a}$ and $\op{n}_b = \opad{b}\op{b}$.
Identical bosonic particles in two modes can be described by the symmetric spin model and the Schwinger representation given above can be used. When it comes to experiments, however, an a priori symmetry assumption is a strong constraint. Its validity has to be shown experimentally.

\subsubsection{Coherent spin states.} \label{sec.css}

Coherent spin states are the most "classical-like" pure quantum states of a symmetric ensemble of $N$ elementary $\sigma=1/2$ spins or of $N$ two-mode Bosons~\cite{Arecchi1972, Radcliffe1971}. They are constructed by placing all $N$ particles in the same single particle state, in any superposition of the two modes 
\begin{equation}
	\ket{\theta, \varphi} = \frac{1}{\sqrt{N!}} [\sin(\theta/2)\e^{-i \varphi/2} \opad{a} + \cos(\theta/2)\e^{i \varphi/2} \opad{b}]^{N} \ket{\mathrm{vac}} 
\end{equation}
where $\ket{{\rm vac}}$ is the vacuum state. 
Any coherent spin state $\ket{\theta, \varphi}$ can be constructed from the state $\ket{0,0} = \frac{1}{\sqrt{N!}} (\opad{b})^{N} \ket{\mathrm{vac}}$ (all N Bosons in mode $\ket{b}$) by a suitable unitary rotation with the angles $(\theta, \varphi)$~\cite{ Dalton2012, Arecchi1972}.

\begin{figure*}
	[t] 
	\begin{center}
		\includegraphics{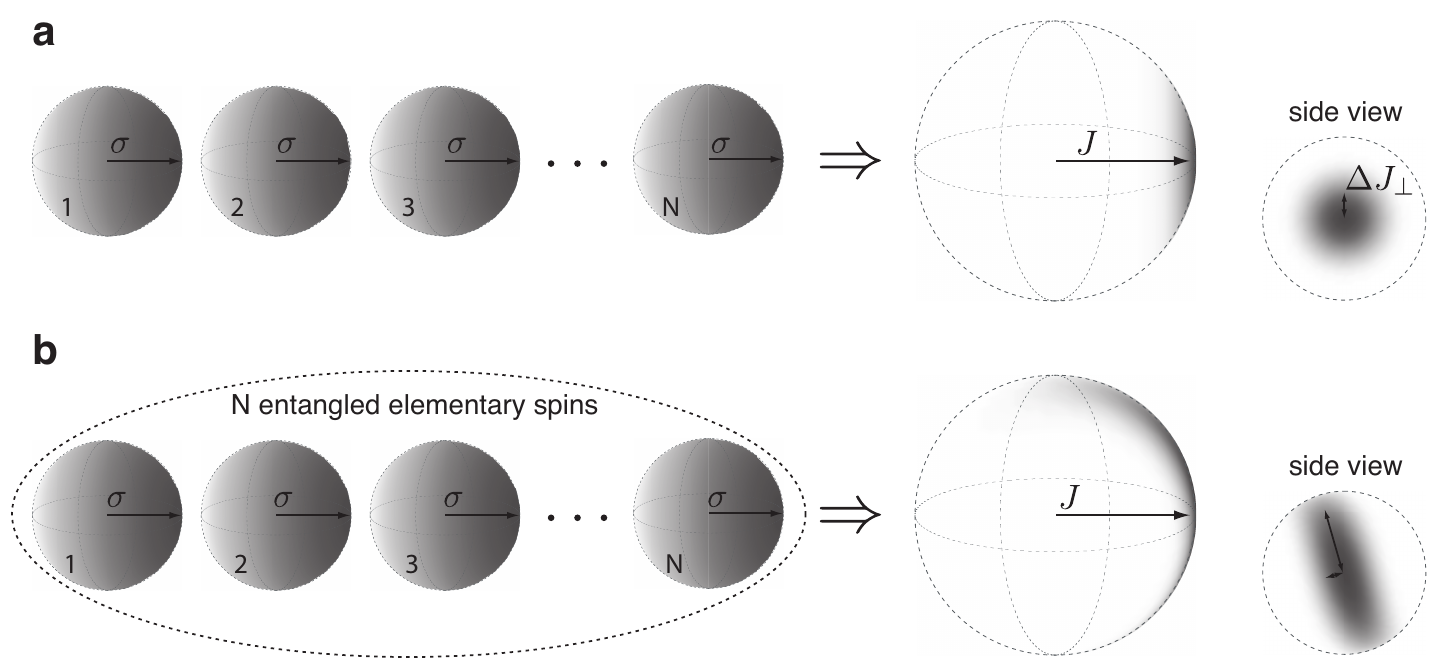} 
	\end{center}
	\caption{Collective spin states composed of elementary spins. {\bf a} Coherent spin states. Coherent spin states in a two-level pseudo spin system can be thought of being composed of identical but otherwise independent elementary spins $\sigma=1/2$. The perpendicular variance $\Delta \sigma_\perp^2=1/4$ of each elementary spin is per definition isotropic. Gaussian addition of independent variances result in the isotropic variance $\Delta J_\perp^2 = N/4$ for the collective spin $J = N/2$. {\bf b} Coherent spin squeezed states. For coherent spin squeezed states the elementary spins have the same isotropic variance. Spin squeezing emerges due to quantum correlations (entanglement) between the elementary spins. The collective spin is squeezed in one direction $\Delta J_{\perp, {\rm min}}^2 < N/4$ and anti-squeezed in the orthogonal one $\Delta J_{\perp, {\rm max}}^2 > N/4$. The Q-representation is used to visualize the spins on the Bloch spheres (see main text).} \label{fig:css} 
\end{figure*}

One important aspect of the coherent spin states is that no quantum correlations between the particles are present.
Therefore, a coherent spin state features equal variance in any direction $J_{\perp}$ orthogonal to the mean spin direction $(\theta, \varphi)$ which, according to the Gaussian addition of variances, is given by the sum of the variances of the $2 J$ elementary spin $1/2$ particles. The perpendicular variances $\D{\op{\sigma}_{\perp}}$ of individual elementary spins are by definition isotropic around $(\theta, \varphi)$ since there are no subsystems that could cause any correlations (figure~\ref{fig:css}a)~\cite{Kitagawa1993}. The Heisenberg uncertainty relation (\ref{eq.spinheisenberg}) for a single elementary spin $\sigma = 1/2$ pointing in $\sigma_{x}$ direction is $\D{\sigma_{z}}\D{\sigma_{y}}=\frac{1}{4} \cdot \frac{1}{4}$ leading to an isotropic variance of 
\begin{equation}
	\D{\Jz}=\D{\Jy} = 2J\cdot\frac{1}{4} = \frac{J}{2} 
\end{equation}
for the collective coherent spin state, which identifies this quantum state as a minimal uncertainty state since $\mean{\Jx} = J$ (see equation~(\ref{eq.spinheisenberg})). We refer to the perpendicular spin fluctuations of a coherent spin state $\D{\op{J}_{\perp}}=J/2 = N/4$ as the shot noise limit.\\

We go back to the first quantization formalism in order to obtain the probability distribution over different sets of basis states -- especially the two Dicke state bases in the directions orthogonal to the mean spin direction.
In order to develop a more detailed understanding of the coherent spin state and its fluctuations we start with the discussion of a special case where each particle is in a $50/50$ superposition of the two modes with $\varphi = 0$ relative phase. Each elementary spin points in $\sigma_{x}$ direction and its quantum state is $\ket{x} = \left(\ket{\frac{1}{2},-\frac{1}{2}}+\ket{\frac{1}{2},+\frac{1}{2}}\right)/\sqrt{2}$, 
where we have chosen the Dicke states in $\sigma_{z}$ direction as the basis states. The probability to observe the elementary spin in state up or down is equal $\left| \bk{\frac{1}{2},\pm\frac{1}{2}}{x} \right|^{2} = 1/2$. \\
The $N$ atom coherent spin state is a collection of these independent elementary spins $\ket{X} = \left[\left(\ket{\frac{1}{2},-\frac{1}{2}}+\ket{\frac{1}{2},+\frac{1}{2}}\right)/\sqrt{2}\right]^{\otimes N}$ and hence the measurement of the $J_{z}$ spin component is equivalent to $N$ measurements on a single spin. The probability distribution over the Dicke states is binomial.
We could have chosen equally the Dicke states in $J_{y}$ direction to describe the spin state which shows again that the spin fluctuations in the directions perpendicular to $J_{x}$, the mean spin direction, are isotropic.\\
A general coherent spin state $\ket{\theta,\vphi}$ in the Dicke basis $\ket{J,m}$ is described by~\cite{ZHANG1990}: 
\begin{equation}
	\ket{\theta,\vphi} = \sum_{m=-J}^{J} \, c_{m}(\theta) \e^{-i (J+m) \vphi} \ket{J,m} \label{eq.cohstate1st} 
\end{equation}
The coefficients $c_{m}(\theta)$ follow a binomial distribution peaked around $\theta$:
\mbox{$c_{m}(\theta) = {2J \choose J + m}^{1/2}\cos(\theta/2)^{J-m} \sin(\theta/2)^{J+m}$}.\\

Due to the fundamental uncertainty in the orthogonal spin directions the mean direction ($\theta,\vphi$) of any spin state can not be defined with arbitrary precision. For a coherent spin state the isotropic angular uncertainty $\Delta \varphi = \Delta \theta$, defined by the ratio of the uncertainty of the perpendicular spin directions $\Delta J_{\perp}$ to the mean spin length $J$, is given by: 
\begin{equation}
	\Delta \varphi = \frac{\Delta \op{J}_{\perp}}{\mean{\op{J}}} = \frac{1}{\sqrt{2J}}=\frac{1}{\sqrt{N}} \label{eq.shotnoise} 
\end{equation}
As outlined above this limit arises as the classical statistical limit in a system consisting of $N$ independent constituents~\cite{Giovannetti2004, Wineland1994}. It is this angular uncertainty which limits the interferometric precision in a Ramsey interferometer.

\subsubsection{Spin squeezed states.}
Quantum correlations between the elementary spin $1/2$ particles of a collective spin $J$ can cause anisotropic fluctuations of the collective spin in the directions perpendicular to the mean spin (figure~\ref{fig:css}b). In reference~\cite{Kitagawa1993} quantum states are considered spin squeezed if the variance of one spin component is smaller than the shot noise limit $J/2$ for a coherent spin state: 
\begin{equation}
	\xin = \frac{2}{J}\D{\op{J}_{\perp{\rm , min}}} < 1 \label{eq.numbersqueezing} 
\end{equation}

Definition (\ref{eq.numbersqueezing}) does not take the second, perpendicular spin direction into account. Due to the Heisenberg uncertainty relation (\ref{eq.spinheisenberg}) reduction of the variance in one direction causes an increase of fluctuations in the other. Real life strategies to obtain spin squeezing might also involve states that are not minimal uncertainty states. For these states, especially for experimentally very important non-pure quantum states, the variance in some other direction than the squeezed direction can be much larger than given by the Heisenberg uncertainty relation. This leads to a reduction of the effective mean spin length $\mean{\vctr{\op{J}}}$. \\
Standard metrologic applications, for example Ramsey interferometry, require a large mean spin length. In order to measure the usefulness of spin squeezed states for these applications another definition of the squeezing parameter was introduced in reference~\cite{Wineland1994} 
\begin{equation}
	\xi_{R}=\sqrt{2J}\frac{\Delta \op{J}_{\perp{\rm , min}}}{\mean{\vctr{\op{J}}}} \label{eq.xir} 
\end{equation}
whose inverse $\xi_{R}^{-1}$ measures the precision gain in a Ramsey interferometric sequence relative to the standard quantum limit (\ref{eq.shotnoise}).

Spin squeezing among $N$ constituents is related to many-body entanglement. In this context a third spin squeezing criterion was defined~\cite{Sørensen2001}: 
\begin{equation}
	\xis=N \frac{\D{\op{J}_{\perp{\rm , min}}}}{\mean{\vctr{\op{J}}}^{2}} = N \frac{\D{\Jz}}{\mean{\Jx}^{2}}  \label{eq.coherentspinsqueezing} 
\end{equation}
Entanglement is detected by the inequality $\xis < 1$ as detailed below. Here we explicitly use the standard assumption throughout this tutorial that the mean spin points in $J_{x}$ direction and the direction of minimal variance -- if not explicitly mentioned -- is the $J_{z}$ direction.\\
$\xi_{S}$ can be used equivalently to $\xi_{R}$ to quantify spin squeezing and precision gain in interferometry and we refer to it as coherent number squeezing or coherent spin squeezing. Both quantities remain a useful parameter to detect entanglement or to measure metrologic precision gain even in a non-symmetric situation. To demonstrate entanglement in a collective spin system without presumptions not only number squeezing is required but also the coherence of the quantum state has to be shown.

\subsubsection{Visualizing spin states: The Husimi Q-representation.}   \label{sec.Qrepresentation}
Employing the Q-representation~\cite{LEE1984}, collective spin states can be conveniently visualized on a generalized Bloch sphere with radius $J$. In order to describe the most general spin state, i.e. pure states and statistical mixtures, the density matrix formalism is used~\cite{Arecchi1972}. The density operator $\op{\rho}$ in coherent spin state basis is given by $\op{\rho} = \int P(\theta,\varphi) \proj{\theta,\varphi}{\theta,\varphi} \mathrm{d}\Omega
$, where the integral covers the full solid angle and ${\rm d}\Omega = \sin(\theta) \,{\rm d}\theta\, {\rm d}\phi$. The quasi-probability distribution $P(\theta,\varphi)$ is normalized to one.
The Q-representation uses the diagonal elements of the density operator to represent the quantum state $Q(\theta,\varphi)=\frac{2J+1}{4\pi}\matel{\theta,\varphi}{\rho}{\theta,\varphi}$. 
It gives the probability of finding the system specified by $\op{\rho}$ in the coherent spin state characterized by $\ket{\theta,\varphi}$.
The interpretation of this representation on generalized Bloch spheres differs from the usual single spin $\sigma=1/2$ Bloch sphere. In that case the dimension of the Hilbert space is two-dimensional and the quantum state representation on the surface of a sphere is exact. However for symmetric collective spin systems the dimension of the Hilbert space is $2J+1$ such that an exact mapping to the surface of a sphere is not possible. The position $(\theta,\phi)$ on the spin $1/2$ Bloch sphere describes the full quantum state, while the position on the generalized Bloch sphere gives only the mean spin direction and, within the following constraints, its fluctuations: 
The Q-representation projects the density matrix on minimal uncertainty states, in particular coherent spin states. The most obvious consequence is that the minimal extension of a quantum state in $(\theta,\vphi)$ on the Bloch sphere is given by the uncertainties of the basis states.\\
The Q-representation for visualizing collective spins is rather easily implemented with modern plotting tools. 
However, there exist more elaborate ways to visualize the quantum state of the two-mode system as described in~\cite{Dalton2012}. 
The Bloch vector visualization described therein has the advantage that the spin noise in all three components of the Bloch vector is directly visible. Also the fragmentation of the quantum state, that is, if the $N$ Bosons do not occupy the same single particle state~\cite{Dalton2012, Leggett2001} can be seen directly by a Bloch vector that lies inside the Bloch sphere.

\subsection{Spin squeezing and entanglement.} 

Entanglement is only a meaningful concept if the subsystems relative to which entanglement is defined are clearly specified~\cite{Amico2008, TerraCunha2008, Goold2009}. Coherent spin squeezing detects entanglement among the elementary spins -- the subsystems are the individual spin $1/2$ particles~\cite{Sørensen2001, Sorensen2001}.  
For $N$ distinguishable particles the definition of a separable state, i.e. non-entangled state, is that its $N$-body density matrix $\rho$ can be written as a direct product of single particle density matrices $\rho^{(i)}$: 
\begin{equation}
	\rho = \sum_{k}p_{k}\rho_{k}^{(1)} \otimes \rho_{k}^{(2)} \otimes \cdots \otimes \rho_{k}^{(N)} \label{eq.separability} 
\end{equation}
$p_{k}$ is a probability distribution to account for incoherent mixtures. Entanglement in many-body systems (for a general review see~\cite{Amico2008,Horodecki2009}) is defined as the non-separability of the density matrix $\rho$ meaning the equality in equation (\ref{eq.separability}) does not hold.\\
Due to technical limitations the individual elementary spins can not be addressed before the measurement in many experiments. However, it is important to note that the elementary spins (the subsystems) have to be in principle distinguishable in order to define entanglement among them in a meaningful way~\cite{Amico2008}. For example, for $N$ particles in a BEC the distinguishability is a priory not given. 
However, by a gedanken local operation one can pinpoint each particle in space at the time of measurement without affecting the spin properties of the system. Distinguishability is now given by the position of each particle. If entanglement is detected in the system, it must have been present in the system before the localization, since local measurements can not generate entanglement~\cite{Plenio2007}. Given that the atoms in the BEC are spaced by more than one wavelength of the detection light (which is usually fulfilled), this gedanken local operation means to overcome the technical limitations for addressability and detection of the individual elementary spins.
The fictitious pinning of the particles at the time of measurement (the photon scattering destroys the BEC) effectively assigns a spatial mode to each particle by which they become distinguishable~\cite{TerraCunha2008, Goold2009}.
The authors of reference~\cite{Benatti2010} show that the notion of entanglement in such a collective spin system can be generalized even without such a fictitious pinning of the individual particles.\\
Not surprisingly different entanglement criteria can be found when regarding the two modes that define the spin $1/2$, as the subsystems~\cite{Milburn2003, Zubairy2006}. It has been argued that this kind of mode entanglement cannot be extracted in experiments without a suitable reference frame~\cite{Dowling2006}. Recently, this problem of a missing reference frame for atomic fields has been addressed experimentally and mode entanglement has been shown in a system of ultracold Bosons~\cite{Gross2011a}.

Without the possibility to address the individual elementary spins before the measurement tests for entanglement, that is, experimentally accessible criteria, based on the collective spin variables are necessary. The observables in most experiments so far are limited to first and second order moments of the distributions functions in different spin directions due to rather small counting statistics and technical noise. Based on these, a complete set of inequalities, that is fulfilled for any separable quantum state, has been found~\cite{Zubairy2006, Toth2007, Toth2009}. Complete in this sense means that assuming the only information available are first $(\mean{\Jx}, \mean{\Jy}, \mean{\Jz})$ and second moments $(\D{\Jx}, \D{\Jy}, \D{\Jz})$ of the distribution functions. Note that these inequalities allow for experimental tests of entanglement, but they are no quantitative measures for entanglement. Such entanglement criteria only allow for the statement: "Yes, that state is entangled with respect to these subsystems." or "No, that state is not entangled with respect to these subsystems." The coherent spin squeezing criterion given in equation (\ref{eq.coherentspinsqueezing}) corresponds to a less general form of such an inequality: all separable states fulfill the inequality $\xis \geq 1$, but a subgroup of entangled states violate it~\cite{Sørensen2001}.
Coherent spin squeezing as an entanglement criterion does not require any symmetry assumption. However, in the fully symmetric subspace many entanglement criteria simplify. For example, the detection of spin fluctuations in one direction below the shot noise limit implies entanglement~\cite{Korbicz2006,Korbicz2005,Korbicz2005a,Wang2003}, that is 
\begin{equation}
	\xin = \frac{4 \, \D{\op{J}_{\perp{\rm , min}}}}{N} = \frac{2 \, \D{\op{J}_{\perp{\rm , min}}}}{J} \geq 1 \label{eq.sqsymmetric} 
\end{equation}
holds for any separable symmetric state. Equation (\ref{eq.sqsymmetric}) is identical to the spin squeezing parameter of Kitagawa and Ueda (\ref{eq.numbersqueezing}) showing that in the symmetric two-mode case entanglement is necessary to redistribute the fluctuations of orthogonal spin components. To avoid misunderstandings about the different definitions of spin squeezing we refer to $\xin$ as number squeezing.

\begin{figure*}[t] 
	\begin{center}
		\includegraphics{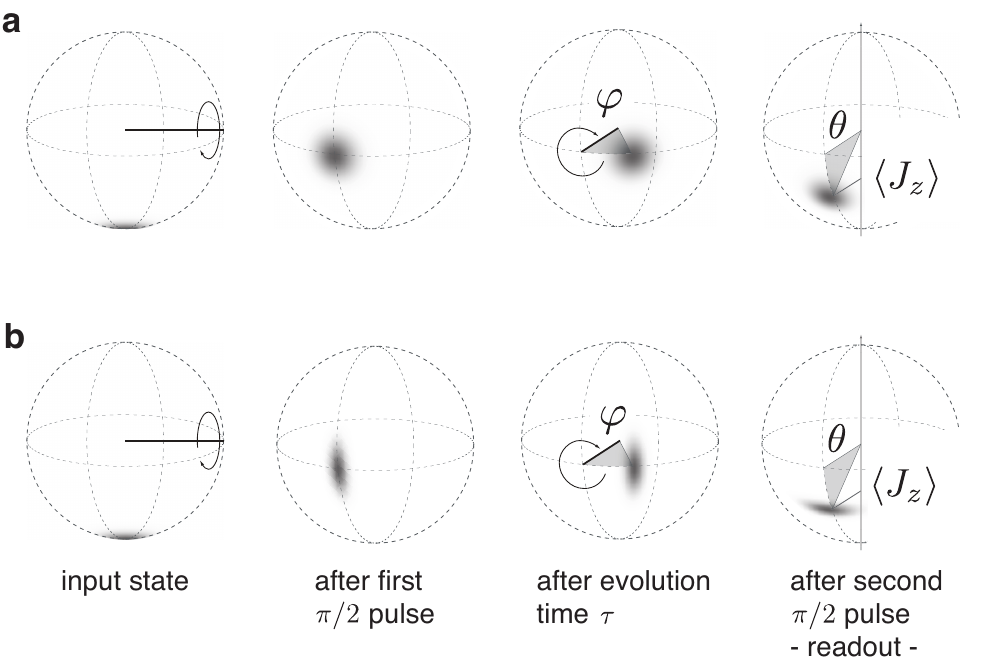} 
	\end{center}
	\caption{Schematic representation of Ramsey interferometry. Panel {\bf a} shows the standard Ramsey protocol depicted on the Bloch sphere. Beam splitters correspond to rotations of the quantum state around an axis in the equatorial plane as indicated by the circular shaped arrows. The sequence is described in detail in the main text. Panel {\bf b} shows a similar protocol using spin squeezed states within the interferometer. The squeezing direction is chosen such that it finally leads to decreased noise in readout direction ($J_z$ direction).} \label{fig:ramsey_bs} 
\end{figure*}

\subsection{Spin squeezing and interferometry.} 

Entanglement in collective spin systems is not only interesting from a conceptual perspective but it has also been shown to provide a useful quantum resource. In 1994 Wineland et al.~\cite{Wineland1994} pointed out, that in particular spin squeezed states can be used to overcome the standard quantum limit in metrology. They can be directly used to boost the precision of conventional Ramsey interferometers. 

Prominent applications of Ramsey interferometry~\cite{RAMSEY1949,RAMSEY1950} are the definition of the time standard~\cite{Santarelli1999}, high precision magnetometry~\cite{Wasilewski2010} or inertia measurements~\cite{Kasevich1992,Gustavson1997}. 
Reviews covering atom interferometry both experimentally and theoretically can be found in references~\cite{Dalton2012, Cronin2009, Reichel2004}.\\
Figure~\ref{fig:ramsey_bs} shows the Ramsey interferometric sequence illustrated on generalized Bloch spheres. The parameter $\delta$ to be measured couples via a $\Jz$ term in the Hamiltonian resulting in a rotation of the spin around the $J_z$ axis during the evolution with rotation frequency $\delta$. The accumulated relative phase $\varphi = \int \delta {\rm d}t$ by which the initial spin state rotated around the $J_z$ direction of the Bloch sphere is finally converted into an observable population difference.

\begin{figure*}[t] 
	\begin{center}
		\includegraphics{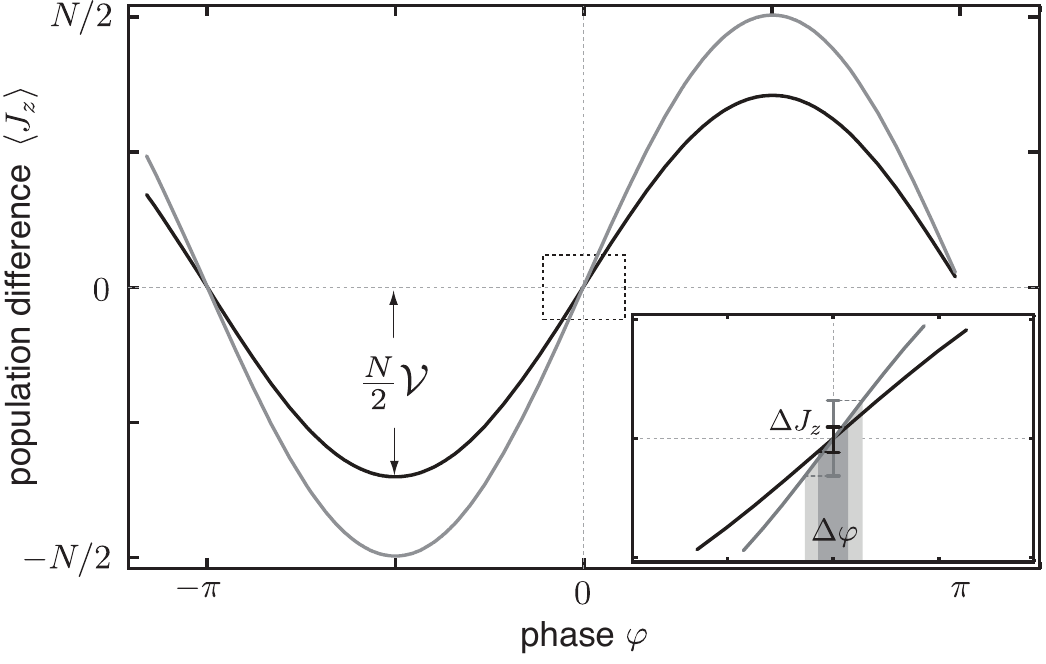} 
	\end{center}
	\caption{Precision limit in Ramsey interferometry. The phase estimation precision in Ramsey interferometry using a coherent spin state (gray) and a spin squeezed state (black) are schematically explained. The main figure shows a Ramsey fringe whose visibility $\mathcal{V}$ is maximal for a coherent spin state ($\mathcal{V}=1$) but smaller for a spin squeezed state. Nevertheless the phase precision for a squeezed state outperforms the precision obtained for classical interferometers as shown in the zoom around the point of highest sensitivity. The projection noise is suppressed for the spin squeezed state such that the ratio of projection noise and slope of the Ramsey fringe is smaller by a factor $\xi_{S}$ compared to the standard quantum limit, which explains the gain in interferometric precision. } \label{fig:ramsey_prec} 
\end{figure*}

The phase resolution of the interferometer
\begin{equation}
	\Delta \vphi^{-1} = \left(\frac{\Delta \Jz }{ \frac{
	\partial \mean{\Jz}}{
	\partial \vphi} }\right)^{-1} 
\end{equation}
is determined by the projection noise $\Delta \Jz$ at the time of readout and the local slope of the Ramsey fringe $\partial \mean{\Jz}/\partial \vphi$ (see figure~\ref{fig:ramsey_prec}). The point of maximum sensitivity (taking finite environmental noise into account) is reached where the mean population difference is zero and the slope is maximal $(
\partial \mean{\Jz}/
\partial \vphi)_{\rm max}= \mathcal{V}N/2$. The visibility $\mathcal{V}$ measures the mean spin length $\mean{\op{J}} =\mathcal{V} N/2$. \\
The amount of precision gain (or loss) relative to standard quantum limit is given by $\xi_{R}^{-2}$, or equivalently by $\left( \xis \right)^{-1}$. The measure can be expressed in visibility $\mathcal{V}$ and spin noise in $J_{z}$ direction at readout $\D{\Jz}$ 
\begin{equation}
	\xis = \frac{4\D{\Jz}}{\mathcal{V}^{2} N} .
\end{equation}
The phase uncertainty measured as the root mean square deviation is $\Delta \vphi = \xi_{S} \,\, \frac{1}{\sqrt{N}}$.\\

The phase $\varphi$ appears in two connected contexts in this tutorial. On the one hand, for coherent spin states, it is a variable characterizing the full quantum state of the system (together with $\theta$). On the other hand, it has the meaning of a relative phase between the two modes of the interferometer (or more general between the modes $\ket{a}$ and $\ket{b}$ of the two-mode system). In both cases the phase $\varphi$ can be interpreted as the mean angle of the spin state in longitudinal direction (that is, in the $J_x$, $J_y$ plane) on the Bloch sphere as schematically shown in figure~\ref{fig:ramsey_bs}. The angular uncertainty in this direction finally limiting the precision of a Ramsey interferometers is $\Delta \varphi$ (cf. section~\ref{sec.css}) \footnote{For two-mode systems a third notion of a relative phase via the definition of phase eigenstates exists that will not be used in this tutorial~\cite{Dalton2012, Gati2006a, Pegg1989}.}.

Spin squeezed states feature reduced noise in one of the spin directions but excess noise in another direction can be present either due to a non-Heisenberg limited quantum state or due to an incoherent mixture of several quantum states. The former might limit precision in standard Ramsey interferometry, but specific correlated quantum states enable even enhanced interferometric precision in a generalized interferometer~\cite{Pezze2009}. The latter is easily limiting interferometric precision at a level above the standard quantum limit and experimentally it requires a large effort to prevent decoherence due to technical noise from the environment or due to finite temperature in the system. Large noise -- quantum or classical -- even in a spin direction that is not directly measured has a degrading effect on interferometric precision in a standard interferometer which arises due to the curved surface of the Bloch sphere. As soon as the noise amplitude is large enough such that the area of uncertainty can no longer be approximated by a plane, the mean spin is effectively shortened and the visibility decreases $\mathcal{V}<1$. \\

Entanglement can be used as a quantum resource in a Ramsey interferometric sequence in different ways. In order to increase the phase sensitivity either the slope of the signal $
\partial \mean{\Jz}/\partial \vphi$ has to be increased or the projection noise $\D{\Jz}$ has to be decreased. 
Slope increase can be reached by Schr{\"o}dinger cat type entanglement which involves maximally entangled states that are very fragile to decoherence. Therefore they have been realized so far with very few particles only~\cite{Leibfried2004,Roos2006, Nagata2007}. \\
Spin squeezing aims to decrease the projection noise. This is possible in gradual steps meaning that depending on the amount of spin squeezing the precision is gradually increased. Therefore -- at least for moderate levels of spin squeezing -- these states are less fragile and they have been realized with a large number of particles but only with a relatively small squeezing factor~\cite{Meyer2001, Leroux2010a, Schleier-Smith2010a, Leroux2010, Fernholz2008, Appel2009, Goda2008, Vahlbruch2008, Esteve2008, Gross2010}. Interferometric sensitivity for a coherent spin state and a coherent spin squeezed state is illustrated in figure~\ref{fig:ramsey_prec}. For the coherent spin squeezed state the decreased quantum fluctuations $\Delta \Jz$ reduce the projection noise while the increased fluctuations $\Delta \Jy$ cause a slight decrease of the mean spin length and therefore of the visibility of the Ramsey fringe. Nevertheless, the ratio of projection noise and slope of the Ramsey fringe -- and therefore the phase sensitivity -- is increased.

\subsection{Heisenberg limit in quantum metrology.} The ultimate limit for metrologic precision is the Heisenberg limit~\cite{Giovannetti2006a}, where the phase estimation error $\Delta \vphi$ is given by $\Delta \vphi = \frac{1}{N}$ for $N$ resources used in a single measurement. This fundamental limit can in principle be reached with both approaches mentioned above, Schr\"odinger cat type entanglement or spin squeezing. 

In the context of quantum metrology the Schr\"odinger cat state is sometimes called a NOON state~\cite{Lee2002}. Its name originates from its form in Fock states basis $\ket{NOON} = (\ket{N,O} + \e^{i\varphi_{N}}\ \ket{O,N})/\sqrt{2}$.
It is a coherent superposition of all atoms in state $\ket{a}$ and zero atoms in state $\ket{b}$ and vice versa. In spin representation the NOON state is the superposition of the two maximal Dicke states $\ket{NOON} = (\ket{J,-J} + \e^{i\varphi_{N}}\ket{J,J})/\sqrt{2}$.
The increase of the signal slope for a NOON state is obvious since the phase acquired between the two components $\varphi_{N} = N \vphi$ is $N$ times larger than for a coherent spin state~\cite{Bouyer1997, Dowling1998, Giovannetti2004}. Experimentally it is important to note that the readout of the interferometer can not be realized by measuring $\mean{\Jz}$. The reason is the vanishing mean spin length $\mean{\Jx}$ of this state. It has been shown that the parity of the state is a useful experimental observable to make use of NOON states in interferometry and to reach the Heisenberg limit~\cite{Bollinger1996,Leibfried2004}. 

Further proposals to achieve Heisenberg limited interferometry involve twin-Fock states $\ket{N/2,N/2}$~\cite{Holland1993, Dunningham2002}. The mean spin length for these states is also vanishing such that special readout schemes are required~\cite{Dunningham2004}. Recently such states have been produced experimentally~\cite{Gross2011a, Bucker2011, Lucke2011} and their usefulness for quantum enhanced metrology has been shown~\cite{Lucke2011}.

Spin squeezed states allow for the optimization of interferometry gain demanding a finite mean spin length such that standard interferometric readout can be used. 
The optimum $\xi_{R}$ for a given mean spin length was found numerically in reference~\cite{Sorensen2001}. An experimental protocol to generate spin squeezed states close to the Heisenberg limit was proposed in reference~\cite{Pezze2005, Bodet2010}. The authors show that the Heisenberg limit can be reached asymptotically in the number of atoms.

The precision of Ramsey type interferometric measurements increases with evolution time within the interferometer (cf. section \ref{sec.intro}). The same inter atomic interactions, that can be used to generate entangled states initially and to overcome the standard quantum limit (as shown in this tutorial), however, might spoil the precision of the interferometer if present during this evolution time~\cite{Grond2010}. Hence, interaction control (for example using Feshbach resonances) is crucial for future applications that use standard interferometric readout. 

Recently it has been emphasized that the Fisher information is the most general criterion to measure phase sensitivity~\cite{Pezze2007, Pezze2009} since it saturates the quantum Cramer-Rao bound~\cite{Wootters1981, Braunstein1994, Helstrom1976}. Calculating the Fisher information for a coherent spin state state evolving under the non-linear Hamiltonian $\op{H} = \chi \Jz^{2}$, where $\chi$ parametrizes the nonlinearity, Heisenberg like scaling for the phase precision was found~\cite{Pezze2009} after long evolutions times where the potential precision gain in a standard interferometer was already decreased. The quantum state here is neither necessarily a NOON state nor a coherently spin squeezed state. However standard interferometric readout can not be used to extract the phase information and a new type of Bayesian readout might be employed which was experimentally demonstrated in~\cite{Pezze2007}.

\section{Preparation of atomic spin squeezed states.} 
Strong effort has been put into the experimental generation and detection of spin squeezed states in large collective spin systems. Early experiments detected spin noise beyond the shot noise limit in a cold ensemble of Cesium atoms prepared by quantum state transfer from non-classical light to the atoms~\cite{Hald1999}, or in a Cesium vapor cell at room temperature by quantum non-demolition measurement~\cite{Kuzmich2000}. Indirect measurements of number squeezing of ultracold atoms in optical lattices and double well systems followed, however, the direct detection of the fluctuations was not possible~\cite{Orzel2001, Li2007, Gerbier2006a, Jo2007, Sebby-Strabley2007, Maussang2010}. 
Several experiments generating and detecting coherent spin squeezing of large collective spins in atomic gases were done in last four years~\cite{Esteve2008, Fernholz2008, Appel2009, Schleier-Smith2010a, Leroux2010a, Gross2010, Riedel2010}. 
Full Ramsey interferometers working beyond the standard quantum limit have been realized using coherent spin squeezed states within the interferometer~\cite{Gross2010, Louchet-Chauvet2010, Leroux2010}. 

\subsection{Spin squeezing with Bose-Einstein condensates.} 
Different methods have proven useful to generate spin squeezed states in atomic samples. Many of them involve light, for example by quantum state transfer, quantum non-demolition measurements or cavity induced interactions~\cite{Hammerer2010}. 
In contrast, ultracold quantum gases offer the possibility to directly use the intrinsic interaction between the atoms. These interactions lead to nonlinear terms in the Hamiltonian necessary to generate spin squeezed states~\cite{Sørensen1999, Sørensen2001}. 

\begin{figure}
	[t] 
	\begin{center}
		\includegraphics{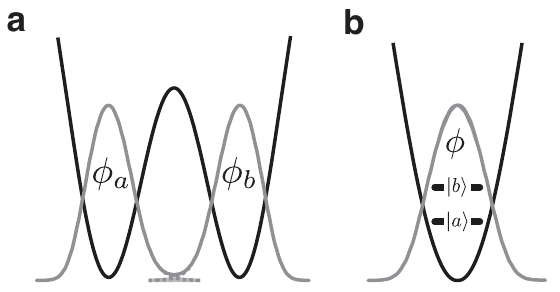} 
	\end{center}
	\caption{Examples for two-mode BECs. {\bf a} A BEC in a double well potential for sufficiently high barrier between the two wells can be approximated as a two-mode system where the mean field modes in the two wells $\phi_a$ and $\phi_b$ define the modes. {\bf b} Another example for a two-mode BEC is a spinor BEC with two populated hyperfine spin states $\ket{a}$ and $\ket{b}$ and identical spatial wave function $\phi$.} \label{fig:example_twomode} 
\end{figure}

We will take intrinsic s-wave interactions among atoms in a two-mode BEC as an example here and discuss how spin squeezing can be generated. Examples for such two-mode systems based on internal or external degrees of freedom are shown in figure~\ref{fig:example_twomode}. 
In such two-mode BECs the inter atomic interactions lead to a nonlinear term in the Hamiltonian proportional to $\Jz^2$~\cite{Milburn1997, Steel1998, Leggett2001}. 
A review on the experimental realization of the external two mode system can be found in~\cite{Gati2006a} and for the internal system details are given in~\cite{Gross2010a, Gross2012}.
The theoretical description for both cases, including the possibility for fragmentation of the quantum state~\cite{Streltsov2007}, has been reviewed in detail recently~\cite{Dalton2012}. Thus, we only sketch the connection between the general Hamiltonian and the pseudo spin description for the external double well case in the following.
The general Hamiltonian is
\begin{eqnarray}
	\op{H} &=& \int \mathrm{d}\vctr{r} \opad{\Psi}\left( -\frac{\hbar \vctr{\nabla}}{2 m} + V_{ext} \right) \op{\Psi} \nonumber \\
	&+& \frac{1}{2} \int \mathrm{d}\vctr{r} \mathrm{d}\vctr{r'} \opad{\Psi} \hat{\Psi}^{' \dagger} V(\vctr{r} - \vctr{r'})\op{\Psi}\op{\Psi}' \label{eq.genHam} 
\end{eqnarray}
from which, by making the two-mode ansatz $\op\Psi(\vctr{r}) = \phi_{a}(\vctr{r}) \op{a} + \phi_{b}(\vctr{r}) \op{b}$, the two-mode Bose-Hubbard Hamiltonian is obtained (for improved readability we did not write the spatial dependence of the operators in equation~(\ref{eq.genHam})  explicitly). 
\begin{equation}
	H_{BH} \approx \frac{\hbar\chi}{4}(\opad{a}\op{a} - \opad{b}\op{b})^{2} - \frac{\hbar\delta}{2} (\opad{a}\op{a} - \opad{b}\op{b}) - \frac{\hbar\Omega}{2} (\opad{a}\op{b} + \opad{b}\op{a}) 
	\label{eq.BoseHubbard} 
\end{equation}
Here we defined three parameters in the Hamiltonian, the nonlinearity $\chi$ characterizing the inter atomic interactions, the Rabi type coupling between the two modes $\Omega$ and a differential energy shift $\delta$.
The remaining step is to write the Hamiltonian~(\ref{eq.BoseHubbard}) in spin language by using the Schwinger representation introduced in equation~(\ref{eq.Schwinger}): 
\begin{equation}
	H = - \hbar\Omega \Jx  + \hbar\chi \Jz^{2} - \hbar\delta \Jz \label{eq.spin} 
\end{equation}

Note that the Hamiltonian in both cases, external and internal degrees of freedom, can be written in the form of equation (\ref{eq.spin}). The actual calculation of the parameters involves spatial integrals over the wave functions and differs in both cases.
In this context it is important to note that the spatial wave functions $\phi$ are not the single particle ground state wave functions in the confining potential, but the mean field wave functions obtained by coupled Gross-Pitaevskii equations for the two modes (For several different approaches see~\cite{Dalton2012, Gati2006a, Streltsov2007, Li2009}). Details on the calculation of the parameters in the Hamiltonian can be found for example in~\cite{Gross2010a, Gross2012, Ananikian2006, Zapata1998}. \\

\begin{figure*}
	[t] 
	\begin{center}
		\includegraphics{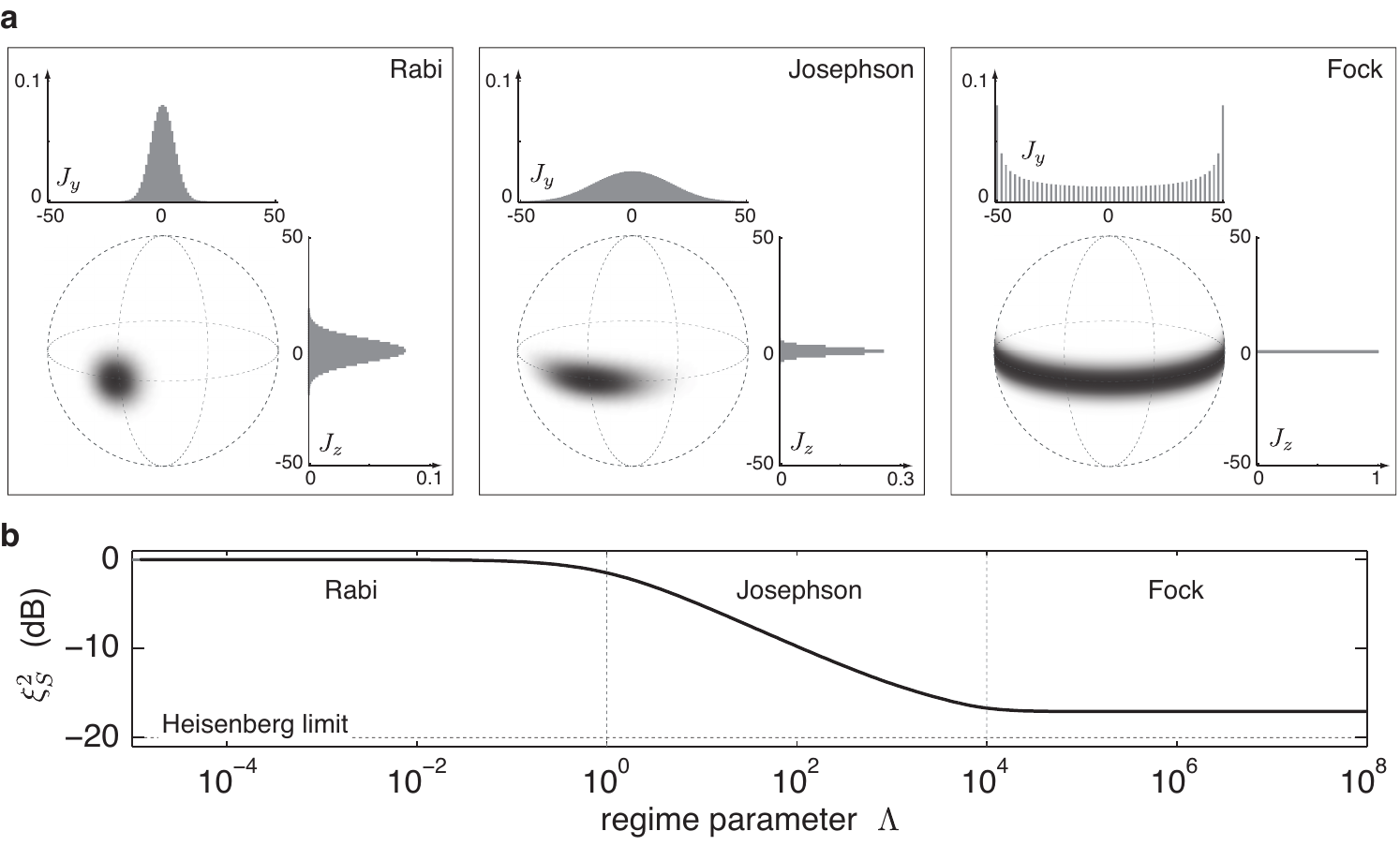} 
	\end{center}
	\caption{Ground state spin fluctuations. Panel {\bf a,} shows the representation of the ground state of the collective spin Hamiltonian~(\ref{eq.spin}) on the Bloch sphere in the three different regimes. The histograms represent the probability distribution of the quantum state over the $J_{z}$ and $J_{y}$ Dicke eigenbasis respectively. The distribution in $J_{z}$ narrows while the variance in $J_{y}$ increases. Note that the distribution of $J_{y}$ in the Fock regime shows fringes with a period of $1/N$ which in principle can be used for increased sensitivity in interferometry~\cite{Pezze2006,Pezze2009}. {\bf b,} shows the coherent spin squeezing parameter $\xis$ as a function of $\Lambda$. The calculation was done for $100$ atoms.} \label{fig:adiabatic_scheme} 
\end{figure*}

Three regimes can be distinguished by looking at the quantum fluctuations in the orthogonal spin directions for the ground state of the Hamiltonian in equation~(\ref{eq.spin}) (see figure~\ref{fig:adiabatic_scheme}). For the following discussion we assume an unbiased situation, that is $\delta = 0$. The important parameter measuring the relative importance of tunnel coupling and interaction is $\Lambda = N \chi/\Omega$. 
This can be seen directly by normalizing the spin operators to spin $1/2$ such that the total number of atoms appears explicitly. Then $\Lambda$ is the ratio of the nonlinear to the linear coefficient. The Rabi regime is the noninteracting limit where $\Lambda \ll 1$. Here the ground state is a coherent spin state with maximal mean spin length $\mean{\Jx} = N/2$ and equal spin fluctuations in the orthogonal directions $\Delta\Jy^2 = \Delta\Jz^2 = N/4$. For increasing interactions fluctuations in the relative atom number (equivalent to fluctuations in $\Jz$) become energetically unfavorable and the system enters the Josephson regime ($1 < \Lambda < N$). Its name originates from the analogy of the Hamiltonian~(\ref{eq.spin}) to the Josephson Hamiltonian in solid state physics~\cite{Leggett2001, Albiez2005, Levy2007}. 
The ground state is a coherent spin squeezed state with reduced spin fluctuations in $J_z$ direction at the cost of increased fluctuations in $J_y$ and hence a reduced mean spin length. The tunnel coupling is negligible in the Fock regime ($\Lambda >> N$). The ground state is an extremely spin squeezed state with vanishing mean spin length. It finally approaches the twin Fock state, an eigenstate of $\Jz$ with eigenvalue zero.

There are two possibilities to control the regime parameter $\Lambda$. For a given total atom number $N$ either the tunnel coupling $\Omega$ or the nonlinearity $\chi$ has to be changed. While the former is straight forward, there are again several possibilities to change the nonlinearity $\chi$. The rather direct approach is to control the scattering properties~\cite{Gross2010}, which is directly possible when using quantum states with accessible Feshbach resonances~\cite{Chin2010}. 
The second method, which has also been realized experimentally, is to control the wave function overlap between the two states~\cite{Li2009, Riedel2010}. 

Assuming the weakly interacting regime can be realized experimentally there are two strategies to generate coherent spin squeezed states. \footnote{Later in this manuscript we will discuss the effects of finite temperature on spin squeezing in a double well setup, where the advantages of starting in a weakly interacting regime are discussed.} The first is an adiabatic approach where the regime parameter is changed slowly driving the system deeper into the Josephson regime~\cite{Pezze2005}. For some experimental settings a fast non-adiabatic approach might be preferable~\cite{Kitagawa1993, Sørensen2001} where spin squeezing develops during the evolution after a quench in $\Lambda$. 
In the following we discuss the advantages and disadvantages of the two strategies.

\subsubsection{The adiabatic approach.} One of the advantages of ultracold atom systems is the ability to control the parameters of the Hamiltonian dynamically. Starting from a strong tunnel coupling the regime parameter $\Lambda$ can be adiabatically changed. Such an adiabatic transformation of the quantum state is a robust way to generate coherent spin squeezed states. The tunnel coupling is present throughout the transformation, making the system less prone to phase noise (fluctuations in $\delta$). This can be intuitively understood by the concept of the spin echo technique~\cite{Vandersypen2004}:  For small differential energy shifts $\delta \ll \Omega$ the continuous tunnel coupling causes a continuous sequence of $\pi$ echo pulses counteracting the dephasing. 
Another advantage of the adiabatic technique is its usefulness in a finite temperature environment, where the initial state is a thermal mixture of the lowest many-body eigenstates of the Hamiltonian (\ref{eq.spin}). In the Josephson regime the adiabatic increase of $\Lambda$ results in an effective ``cooling'' of the system~\cite{Esteve2008}. The term cooling should be used with care here since no entropy is removed from the system. With increasing parameter $\Lambda$ each of the lower eigenstates of the system features reduced fluctuations in $\Jz$, however at the cost of increased fluctuations in $\Jy$ and a loss in coherence. Nevertheless, the gain in $J_z$ direction outweighs the loss of coherence for small enough temperatures, such that the coherent spin squeezed regime can be reached by this method.

It has been theoretically shown that the Heisenberg limit can be reached asymptotically in the number of particles by the adiabatic technique~\cite{Pezze2005, Bodet2010}. 
Experimentally, however, it is a major problem to assure adiabaticity during the full range of the parameter change. At least when changing the regime parameter by tuning the tunneling coupling (for fixed and typically small nonlinearity $\chi$) the required timescale for adiabaticity eventually diverges setting a limit to the attainable coherent spin squeezing factor $\xi_S$. The long times required to generate coherent spin squeezing adiabatically are the main disadvantages of this approach. 

\subsubsection{The diabatic approach.} For some experimental systems the robust but slow adiabatic approach to generate spin squeezing might not be optimal. A diabatic technique, in which the Hamiltonian parameters are quenched (changed non-adiabatically), can produce spin squeezing in a much shorter time. The idea of the diabatic one-axis twisting was introduced in 1993 by Kitagawa and Ueda. Starting with a coherent spin state with mean spin in $J_x$ direction the tunnel coupling is suddenly switched off such that the nonlinear term in the Hamiltonian governs the evolution $\op{U}(t) = \e^{-i \,t\, \chi \Jz^{2} }$~\cite{Kitagawa1993, Sørensen2001, Li2009}. The nonlinearity causes a shearing of the uncertainty region on the Bloch sphere since the rotation speed around $J_z$ is itself $J_z$ dependent. The scheme is similar to Kerr effect based squeezing protocols in quantum optics where a material with a finite Kerr nonlinearity is used such that the refractive index $n_{\mathrm{light}} \propto n_{2} \abs{E}^{2}$ is proportional to the light intensity $\abs{E}^{2}$. The light experiences intensity dependent phase modulation within this medium resulting in quadrature squeezing~\cite{Bachor2004}.

\begin{figure*}
	[t] 
	\begin{center}
		\includegraphics{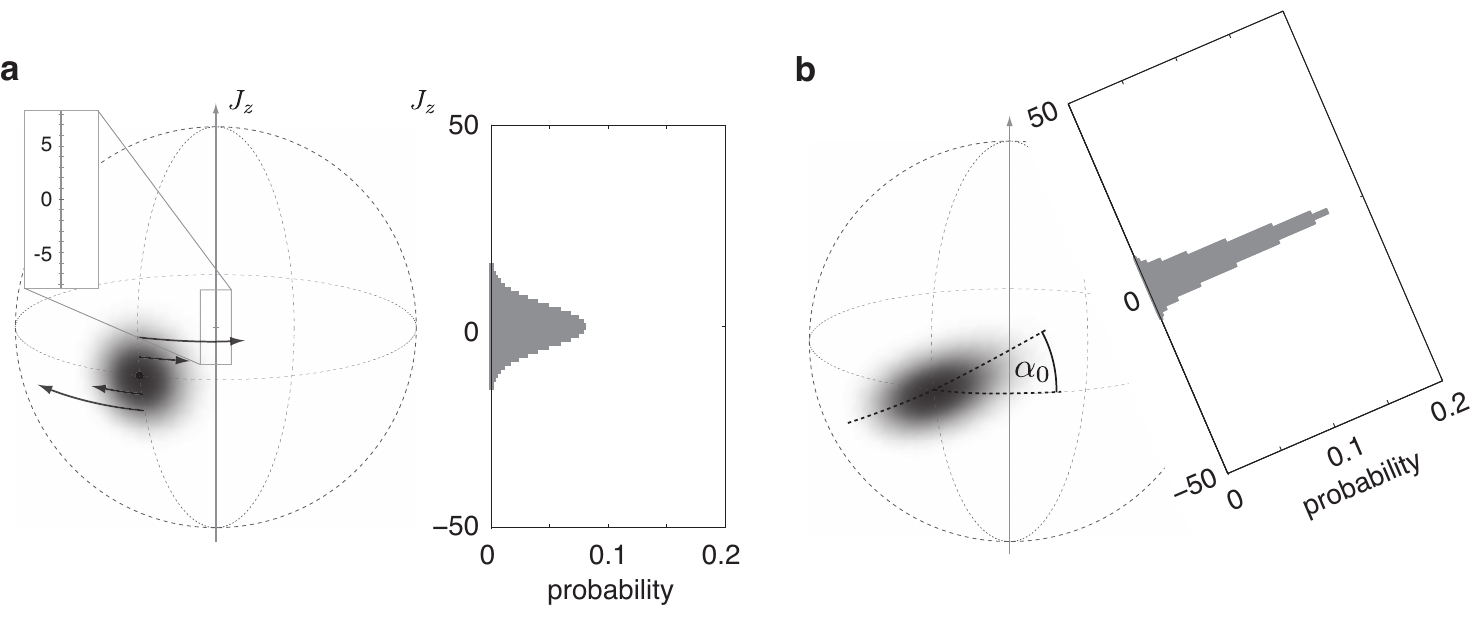} 
	\end{center}
	\caption{Non-adiabatic spin squeezing by one axis twisting. Panel {\bf a} illustrates the initial coherent spin state on the equator of the Bloch sphere. The histogram shows the binomial probability distribution over the Dicke states. The zoom window illustrates the quantization on the $J_{z}$ axis -- the different Dicke states -- of which each eigenstate $m$ rotates with a different angular frequency $m  \chi$ around the vertical axis.  Panel {\bf b} illustrates the quantum state after a short evolution time $\tau < \tau_{\mathrm{deph}}$. The isotropic uncertainty developed into an elliptical one with spin squeezing present under an angle $\alpha_{0}$. The histogram shows the squeezed probability distribution over the eigenstates in a coordinate system rotated by $\alpha_{0}$. } \label{fig:diabatic_scheme} 
\end{figure*}

The one axis twisting evolution of an initial coherent spin state can be nicely visualized. Earlier in this manuscript it has been discussed how a coherent spin state on the equator of the Bloch sphere $\ket{\theta=\pi/2,\vphi=0}$ can be described as a coherent superposition of several Dicke states where the probability distribution over these basis states is binomial. Each Dicke state $\ket{J,m}$ composing the coherent spin state rotates with a different frequency around the $J_{z}$ axis where the difference in rotation frequency between next neighboring Dicke states is $\chi$ (figure~\ref{fig:diabatic_scheme}a). 
For short evolution times $\tau < \tau_{\mathrm{deph}}$ this shearing effect results in coherent spin squeezing under an axis rotated by the angle $\alpha_{0}(\tau)$ with respect to the equator of the Bloch sphere (figure~\ref{fig:diabatic_scheme}b). $\tau_{\mathrm{deph}} = (\sigma_{m} \chi)^{-1}$ is the dephasing time, after which the mean spin length \mbox{$2 \mean{\op{J}}/N $ } has dropped from a value close to unity to $\e^{-1}$ due to the interaction induced spread of the state around the Bloch sphere~\cite{Wright1996}. In general $\tau_{\mathrm{deph}}$ is inversely proportional to the extension of the quantum state over the Dicke basis $\sigma_{m}$ which, for a coherent spin state on the equator of the Bloch sphere, is $\sigma_{m}=\sqrt{N}/2$. \\
At evolution times $\tau \gtrsim \tau_{\mathrm{deph}}$ reduced quantum fluctuations under a certain axis are still present but the mean spin length is strongly reduced such that the quantum state is no longer optimally coherently spin squeezed. After even longer times the dynamics show highly non-classical interference effects~\cite{Yurke1986a, Greiner2002, Ferrini2008} finally resulting in a revival of the mean spin length at $\tau_{\mathrm{rev}} = 2\pi\,\chi^{-1}$ when each neighboring pair of Dicke states is in phase again~\cite{Greiner2002}.
The best achievable noise suppression in general interferometry still increases with evolution time even if coherent spin squeezing degrades. It has been recently shown~\cite{Pezze2009} that a new type of Bayesian interferometer readout can be employed to make use of these quantum states. 
As typical for macroscopic superposition states the highly entangled states occurring for long evolution times are extremely sensitive to decoherence. For example, particle loss leads to an exponential decay of the revival contrast with a $N$ times enhanced decay constant~\cite{Sinatra1998}.

The maximum squeezing achievable with the diabatic one axis twisting technique is $\xis \approx N^{-2/3}$~\cite{Sørensen2001, Kitagawa1993}.
This has to be contrasted to the maximum squeezing that can be generated using the adiabatic technique which approaches the Heisenberg limit $\xis \approx  N^{-1}$ at the boundary to the Fock regime where the mean spin length is still reasonable high~\cite{Pezze2005}. The maximum squeezing is better in the adiabatic case however the adiabaticity criterion requires long evolution times $\tau_{\mathrm{adiab}} \approx 1/\chi$. The best squeezing in the one axis twisting protocol is achieved after $\tau_{dia} \approx N^{-2/3} / \chi$, a factor of $N^{2/3}$ faster favoring the diabatic protocol for certain applications.

The generation of spin squeezing might be optimized using different Hamiltonians. Stronger maximal squeezing, for example, can be reached by a two axis twisting scheme~\cite{Kitagawa1993}. 
Another example, optimizing the required time to achieve squeezing, is to use evolution under the Hamiltonian (\ref{eq.spin}) after a quench to a small but finite tunnel coupling (into the Josephson regime)~\cite{Zibold2010}. 
For initial states with relative phase $\varphi = \pi$ the evolution leads first to coherent spin squeezing and for longer times to Schr\"odinger cat like states~\cite{Micheli2003, Lapert2011}. 

When discussing the maximum achievable coherent spin squeezing one has to keep in mind that with decreasing $\xis$ the generated states become more and more sensitive to decoherence. For the exemplary case of the one axis twisting evolution this problem has been studied in detail taking into account particle loss, random dephasing (for example due to noise in $\delta$, see equation (\ref{eq.spin})) and multi mode effects due to finite temperature which are always present in real condensates~\cite{Li2008, Sinatra2011, Sinatra2012}. Including these effects the achievable coherent spin squeezing is no longer proportional to $N^{-2/3}$ but tends to a finite but small constant value in the thermodynamic limit $N \rightarrow \infty$, showing that a large metrology gain can still be reached~\cite{Sinatra2011}. Reference~\cite{Sinatra2012} gives a review on decoherence effects in dynamically generated spin squeezing with BECs and points out the effective dephasing effect of the different sources of decoherence.

\begin{figure*}
	[t] 
	\begin{center}
		\includegraphics{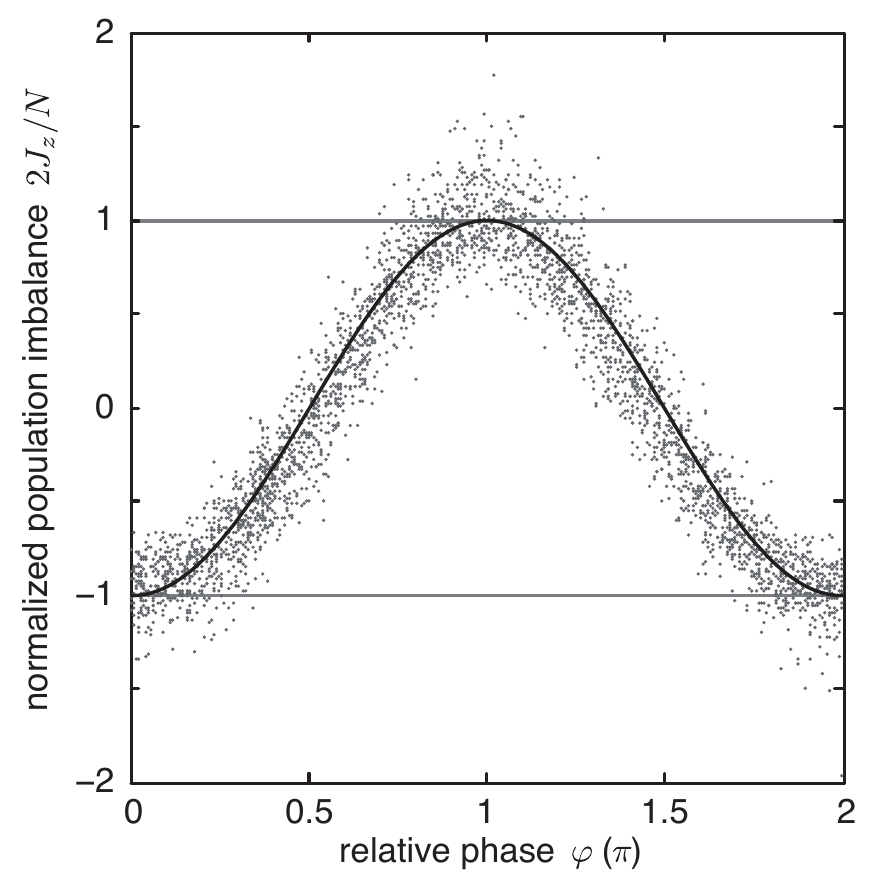} 
	\end{center}
	\caption{Ramsey fringe using a coherent spin squeezed state. The figure shows a Ramsey fringe obtained for a coherent spin squeezed state prepared using the one axis twisting sequence for an evolution time $\tau \ll \tau_{\rm deph}$. The solid line is a sinusoidal fit through the data from which the visibility (here $\mathcal{V} = 1.00\pm0.02$) and mean spin length $\mean{\op{J}} = N \mathcal{V}/2$ can be deduced. This figure was originally published in the supplementary online material to~\cite{Gross2010}.} \label{fig_ramseyfringe} 
\end{figure*}

\section{Direct observation of coherent spin squeezing -- experimental challenges.} Both approaches the adiabatic and diabatic one outlined in the last section require precise control over the experimental parameters. Especially so far neglected uncontrolled differential energy shifts $\delta$ are the main experimental challenge to observe spin squeezing. Another important issue, mainly in experiments with BECs in double well traps, is finite temperature which easily limits the observable spin squeezing. 
The detection of atomic noise at the shot noise level in a small optically thick cloud is challenging and requires careful calibration of the imaging system. Furthermore, a method to detect the mean spin length has to be available experimentally in order to detect coherent spin squeezing.
In this section we will discuss these challenges in detail, starting with the detection issues.

\subsection{Measurement of the mean spin length.}  
When working with internal degrees of freedom the measurement of the mean spin length can be done by a Ramsey experiment~\cite{RAMSEY1949, RAMSEY1950}. 
A Ramsey scheme is usually relatively easy to implement given the availability of precise microwave and radio frequency sources (or Raman lasers) to control the tunnel coupling~\cite{Vandersypen2004}. This coupling $\Omega$ usually can be made much stronger than the nonlinearity $\chi$ such that the Rabi regime is reached. When working with two photon transitions, which is often required to use states less sensitive to magnetic field fluctuations, care has to be taken to accurately determine the resonance. The resonance is shifted during the coupling with respect to the free evolution times due to the ac-Stark shift induced by the coupling fields~\cite{Gross2010a, Gross2012}.
The readout of the Ramsey interferometer requires precise detection of the atom number difference in the two internal states over the full dynamic range. The mean spin length $\mean{\op{J}} = N\mathcal{V}/2$ is then inferred from the visibility $\mathcal{V}$ of the Ramsey signal. An example for such a Ramsey fringe revealing a visibility of $\mathcal{V}=1.00\pm0.02$ is shown in figure~\ref{fig_ramseyfringe}~\cite{Gross2010}. 

\begin{figure*}
	[t] 
	\begin{center}
		\includegraphics{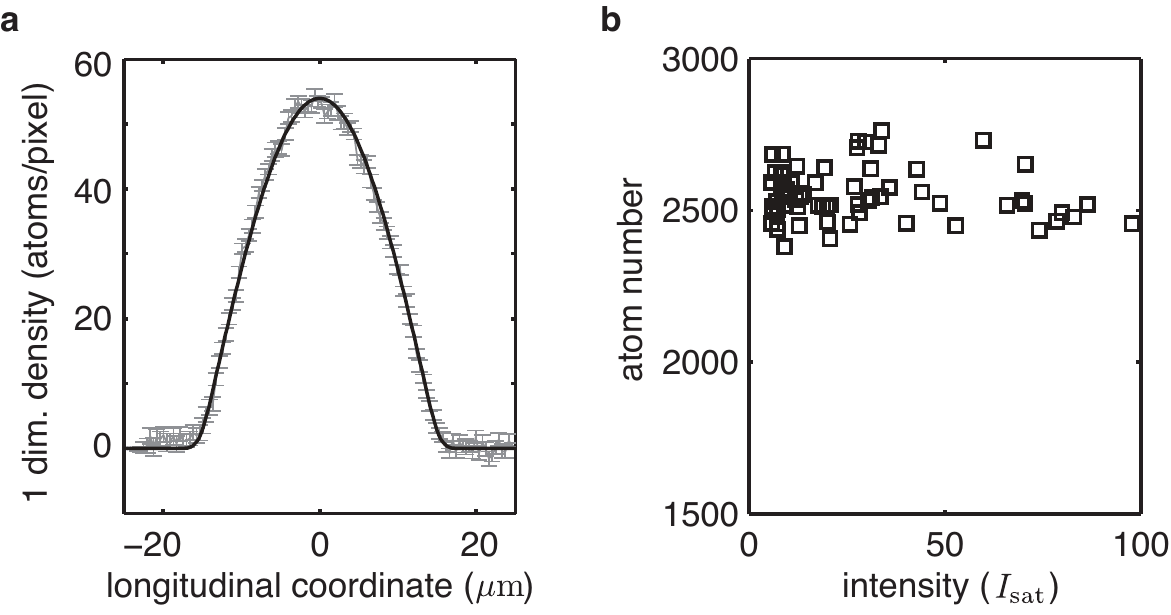} 
	\end{center}
	\caption{Calibration of the imaging system. {\bf a} The correction factor $c_{{\rm gpe}}$ is determined by best matching the observed profile to the numerical simulations. The figure shows an exemplary calibration.  {\bf b} The Clebsch-Gordan coefficient is obtained in a second step by requiring imaging independent atom number measurement. The figure shows an example of a successful calibration.} \label{fig:calibration} 
\end{figure*}

For external degrees of freedom, for example in a double well trap for BECs the mean spin length can be revealed from the interference pattern after time of flight expansion~\cite{Ketterle1999a, Albiez2005, Gati2006, Esteve2008}. 
Additional complications might arise from a finite imaging resolution or due to excitations of additional spatial modes when expanding the condensate, both of which limit the visibility of the interference pattern in a single shot. Care has to be taken when extracting the average visibility $\mathcal{V}$. In general the analysis method chosen has to include the effects of the possibly reduced single shot visibility, which might also be caused by multi mode effects in the initial double well~\cite{Gross2010a, Gross2012}.
Exemplary interference measurements for a BEC released from a double well trap are shown in figure~\ref{fig_numberphase}b later in this manuscript.

\subsection{Absolute atom number measurements at the shot noise level.} 
The detection of number squeezing requires an accurate calibration of the deduced total atom number and the linearity of the imaging system. 
Simultaneously a high spatial resolution is often required~\cite{Esteve2008, Gross2010}. 
To realize sufficiently large nonlinearities $\chi$ in the mesoscopic atom number regime a high atomic density and therefore tight confinement of the atoms is necessary, leading to optically think samples.
This complicates the use of in principle single atom sensitive fluorescence detection methods~\cite{Bakr2009,Sherson:2010} which suffer from light induced collision losses if the individual atoms are not spatially separated during the detection. 
An alternative is to use absorption imaging~\cite{Ketterle1999a} in the high saturation intensity regime~\cite{Reinaudi2007} which provides good signal to noise ratio and high spatial resolution, however, its calibration is non trivial as outlined below.

\subsubsection{Calibration of the imaging system.}
In high intensity absorption imaging~\cite{Reinaudi2007} the full Beer-Lambert absorption formula is necessary to calculate the atomic column density $n_{i,j}$ from the counts per pixel $N_{\gamma{\rm ,pic,i,j}}$ on the picture containing the absorption information and on the reference picture $N_{\gamma{\rm ,ref,i,j}}$ (\mbox{${\rm i,j}$ indexes} the CCD camera pixels but it is mostly omitted below to ensure better readability). 
For the image analysis the mean light intensity on the reference picture has to be normalized to the mean light intensity on the picture with the atomic signal. For its determination the area containing the atomic signal is not taken into account.
Not only the optical density $O_{d}=\ln(N_{\gamma{\rm ,ref}}/N_{\gamma{\rm ,pic}})$ but also the difference in the counts $\Delta = (N_{\gamma{\rm ,ref}} - N_{\gamma{\rm ,pic}})$ contribute to the signal: 
\begin{equation}
	n = \frac{d^{2}}{\sigma_{0}} ( \frac{1}{c_{cg}} \, O_{d} + \frac{c_{{\rm ccd}} \, c_{{\rm gpe}}}{\tau} \, \Delta) \label{eq.imaging} 
\end{equation}
Here $d$ is the linear extension of the CCD pixel taking the magnification into account, \mbox{$\sigma_{0}=3 \lambda^{2}/2\pi$} the resonant cross section, $c_{cg}$ the Clebsch-Gordan coefficient, $c_{{\rm gpe}}$ a correction factor obtained from a comparison to simulations as explained below and $\tau$ is the imaging pulse duration. The factor $c_{{\rm ccd}} = \hbar \omega / (d^{2} \eta Q I_{sat})$ contributes to the linear part of the formula where $Q$ is the total quantum efficiency and $\eta$ the gain factor of the camera. $\omega$ is the light angular frequency and $I_{sat}$ the saturation energy of the transition.
The gain factor $\eta$ is determined by measuring the noise curve $\Delta N_\gamma(\langle N_\gamma \rangle)$ of the CCD camera. 
It can be measured over the full dynamic range using an incoherent light source (e.g. a LED) such that the light hitting the active region of the CCD is uncorrelated featuring shot noise limited noise characteristics. The factor $\eta$ follows from the slope of the CCD's noise curve via a linear fit in the working region where the camera noise is not read out noise dominated. 

The calibration of the $c_{{\rm ccd}}$ factor is based on the quantum efficiency measurement which itself relies on the power meter calibration. Furthermore, for a typical numerical aperture around $0.5$ approximately $5\%$ of the solid angle is covered by the imaging lens such that a non negligible fraction of the scattered photons reaches the camera. This effect is not taken into account in equation (\ref{eq.imaging}).
A possible method to cross check the inferred number of atoms is to prepare the BEC in a very well known trap configuration and image it with very high imaging intensity (typically $50\, I_{sat}$). In this regime only the linear part of the Beer-Lambert formula contributes, but the signal to noise ratio is poor. Imaging is done with a very short pulse ($2\,\mu$s) to minimize diffusive broadening of the profile due to photon scattering. The extracted average profile along an axis much bigger than the optical resolution is calculated for different correction factors $c_{{\rm gpe}}$ and compared  to theoretical profiles obtained from three dimensional Gross-Pitaevskii equation simulations varying the total numbers of atoms (figure~\ref{fig:calibration}a). The correction factor is finally found by minimizing the quadratic deviation between the simulated and measured profile. The expected corrections are small such that $c_{{\rm gpe}}$ is should be close to unity. 

\begin{figure}
	[t] 
	\begin{center}
		\includegraphics{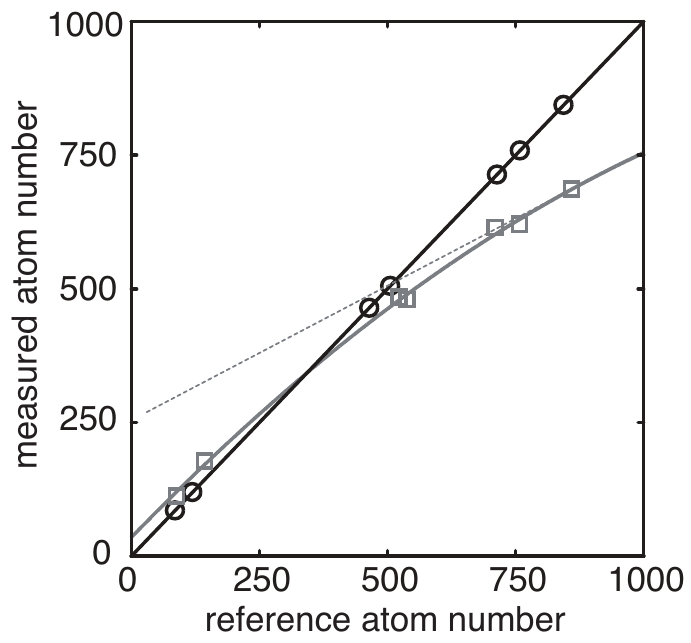} 
	\end{center}
	\caption{Nonlinear effects in absorption imaging of small atomic clouds. For small clouds (gray boxes) and up to intermediate imaging intensities (in the optimal signal to noise regime) the atom number is systematically underestimated with rising optical density. Linearizing the result around $750$ atoms yields a factor of two smaller slope  while the absolute number is underestimated by only $15\%$. This results in a strong bias of the measured atom number fluctuations in this regime. The reference atom number in this figure was measured for $400\,\mu$s expansion time and with high imaging intensity $I/I_{\rm sat}$ (black circles)~\cite{Esteve2008, Gross2010a, Gross2012}. This figure was originally published in the supplementary online material to~\cite{Esteve2008}.} \label{fig:nonlinear_effects} 
\end{figure}

Calibration of the imaging is completed by the measurement of the Clebsch-Gordan coefficient $c_{cg}$. This is done by imaging the atoms with varying imaging intensities $I/I_{sat}$. A correct Clebsch-Gordan factor means intensity independent atom number measurements (see figure \ref{fig:calibration}b). \\
The total atom number follows from direct summation over the region where the atoms are detected (not by a fit which might bias fluctuation measurements). The size of this region is typically chosen to three standard deviations as obtained from a Gaussian fit to the atomic cloud.

\begin{figure}
	[t] 
	\begin{center}
		\includegraphics{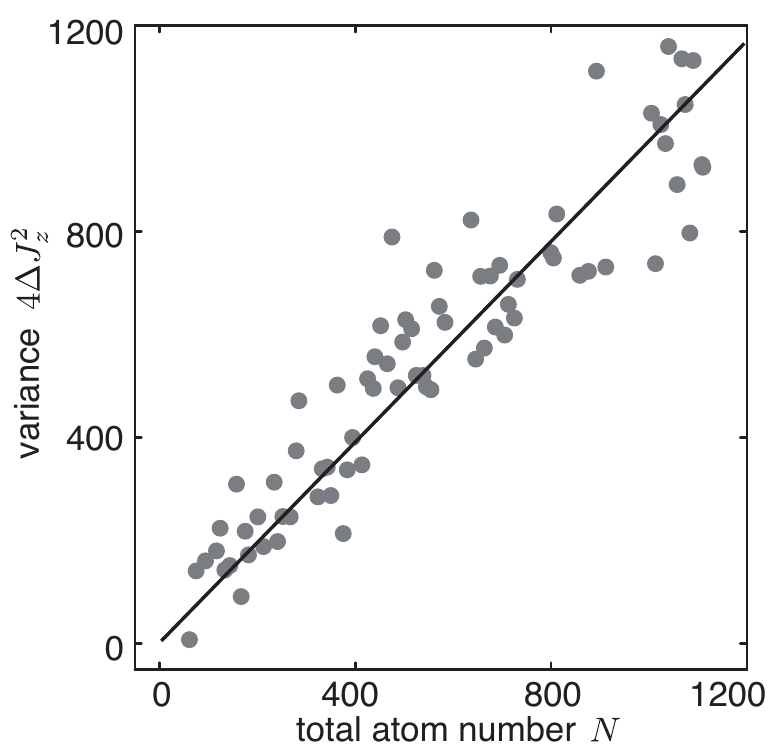} 
	\end{center}
	\caption{Correctly calibrated imaging system. The variance of a coherent spin state on the equator of the Bloch sphere is measured versus the total atom number to test the imaging calibration. The linearity (with a fitted slope close unity, black line) of the measured data confirms the independently obtained calibration of the imaging system. This figure was originally published in the supplementary online material to~\cite{Gross2010}.} \label{fig_imagingLinearity} 
\end{figure}

\begin{figure*}
	[t] 
	\begin{center}
		\includegraphics{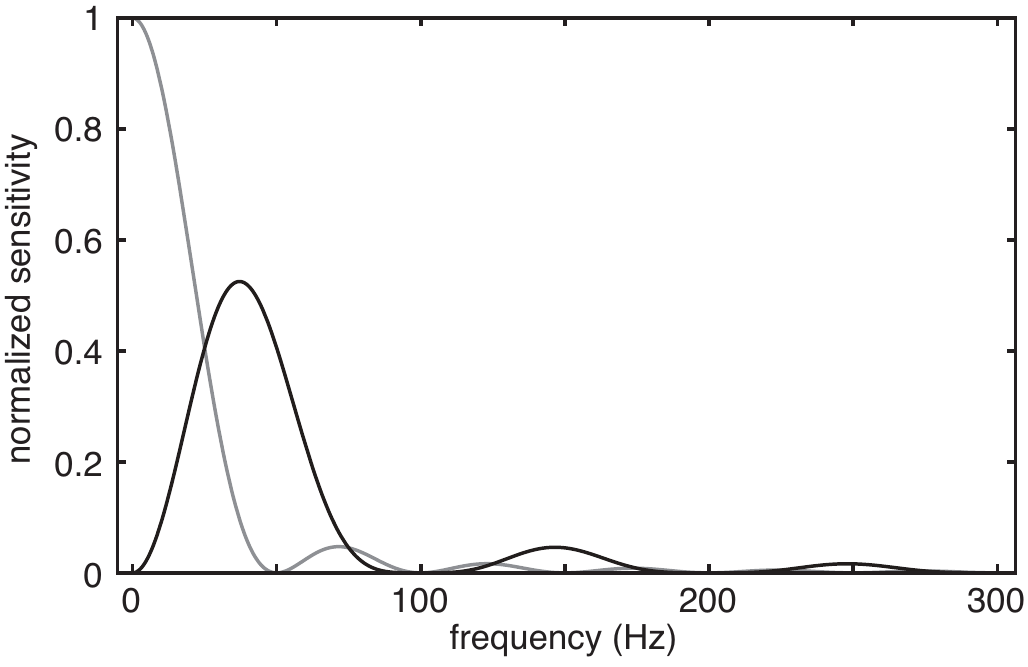} 
	\end{center}
	\caption{Spectral sensitivity to phase noise. The figure shows the normalized spectral phase noise sensitivity of a Ramsey sequence for an exemplary free evolution time of $20\,$ms (gray line). The black line shows the altered sensitivity including a spin echo after $10\,$ms.} \label{fig_phasenoise} 
\end{figure*}

\subsubsection{Non-linear effects.}   
Quantitative measurements of the atom number using absorption imaging is limited to atomic clouds with a size larger than the optical resolution of the imaging system. For to small clouds the non-linear part of equation (\ref{eq.imaging}) causes a systematic error in the detected atom number. This can be easily understood in the limit where the optical resolution is much larger than the pixel size. The detected column density $n$ per pixel is calculated from photon counts integrated over the size of the pixel 
\begin{equation}
	n \propto \ln \left( \frac{\int_{\rm pix} N_{\gamma{\rm ,ref}}}{ \int_{\rm pix} N_{\gamma{\rm ,pic}}} \right) 
\end{equation}
where we omitted the linear term in equation (\ref{eq.imaging}). Correct atom number calculation $n_{\rm true}$ however requires to calculate 
\begin{equation}
	n_{\rm true} \propto \int_{\rm pix} \ln \left( \frac{ N_{\gamma{\rm ,ref}}}{ N_{\gamma{\rm ,pic}}} \right) 
\end{equation}
In general $n \neq n_{\rm true}$ holds such that the calculated atom number is wrong. When the pixel size is smaller than the point spread function $f$ of the imaging system the same argument holds, but the main averaging effect is due to the convolution with the point spread function, e.g. the replacement $N_{\gamma} \rightarrow (N_{\gamma} \ast f)$ has to be made, where $(g \ast f)$ means the convolution of the functions $f$ and $g$. Figure~\ref{fig:nonlinear_effects} shows an example of this effect. The nonlinearity in the atom number calculation for high optical densities and small atomic clouds is clearly visible. It leads not only to an underestimation of the absolute atom number but also to an underestimation of the fluctuations in the high density region due to the smaller local slope. By slightly expanding the atomic clouds before imaging this effect can be minimized.

\subsubsection{Photon shot noise.} 
The atoms are detected via their resonant interaction with the probe light whose noise characteristics add to the atomic noise of interest. At least the shot noise of the detection light adds to the detected noise, but especially in absorption imaging interference fringes easily increase this minimal background noise level. These fringes originate from motion of the imaging systems optics or air motion in the light path. Special care has to be taken to eliminate these effects. 
The detection light shot noise contribution can be inferred pixel by pixel on both, the reference and absorption picture, using the camera noise calibration curve. 
Standard error propagation of equation~(\ref{eq.imaging}) allows for a conversion of the photonic variance $\D{N_{\gamma {\rm , pic (ref)}}}$ into atomic variance: 
\begin{eqnarray}
	\Delta N_{\rm psn,i,j}^{2}= \frac{d^{4}}{\sigma_{0}^{2}} \, &\Bigg\{& \frac{1}{c_{cg}^{2}} \left( \frac{ \D{N_{\gamma {\rm , pic}}} }{ N_{\gamma{\rm ,pic}}^{2} } + \frac{ \D{N_{\gamma {\rm , ref}}}}{ N_{\gamma{\rm ,ref}}^{2} } \right)  \\
	&+& \left(\frac{c_{{\rm ccd}} \, c_{{\rm gpe}}}{\tau}\right)^{2} \left( \D{N_{\gamma {\rm , pic }}} + \D{N_{\gamma {\rm , ref}}} \right) \nonumber \\
	&+& \frac{2 \, c_{{\rm ccd}} \, c_{{\rm gpe}}}{c_{cg}\, \tau} \left( \frac{ \D{N_{\gamma {\rm , pic}}} }{ N_{\gamma{\rm ,pic}} } + \frac{ \D{N_{\gamma {\rm , ref}}} }{ N_{\gamma{\rm ,ref}} } \right) \Bigg\} \nonumber
\end{eqnarray}
Here the photon shot noise contribution $\Delta N_{\rm psn,i,j}^{2}$ is calculated by a expansion up to second order, assuming a Gaussian distribution for the photon statistics $p(N_{\gamma})$. Photon shot noise on different pixels (i,j) is uncorrelated such that the total contribution in a given area of the CCD chip $\D{N_{{\rm psn}}}$ can be obtained by summation over the variance per pixel. The photon shot noise is independent of the atomic noise $\D{N}$. Thus, the best estimate of the true atomic fluctuations can be obtained by subtracting it from the detected fluctuations $\D{N} = \D{N_{{\rm det}}} - \D{N_{{\rm psn}}}$. The uncertainty in the estimated amount of photon shot noise is typically in the order of a few percent.\\
In the strong saturation regime the imaging light intensity $I$ controls the transparency of the atomic cloud~\cite{Reinaudi2007}. The imaging signal to noise ratio can be optimized by choosing the proper light intensity depending on the atomic density and the camera parameters.

\subsubsection{Independent tests of the imaging calibration.}
From the discussion above it is clear that correct calibration of the imaging system is non-trivial. Therefore it is important to have an independent test of the deduced atomic variances. 
One possibility is to hold an atomic sample in the trap until a large fraction of the atoms is lost by background gas collisions. This uncorrelated one-body loss tends to restore a Poissonian atom number distribution. However, the Poissonian limit is only reached asymptotically and the fluctuations might be super-Poissonian for an initial distribution governed by experimental instabilities or slightly sub-Poissonian if the loss is dominated by two or three body collisions~\cite{Esteve2008, Gross2010a, Gross2012, Itah2010, Whitlock2010}. \\
When working with internal degrees of freedom coherent spin states can be prepared by precise electromagnetic coupling pulses. Their well known noise characteristics, that is, relative atom number fluctuations at the shot noise limit, can be directly used for the imaging calibration. Figure~\ref{fig_imagingLinearity} shows such a calibration check which reveals a correctly calibrated imaging over a large range of total atom numbers~\cite{Gross2010, Gross2010a, Gross2012}.

\subsection{Differential energy shifts.} In the discussion of the squeezing generation we assumed vanishing external noise during the evolution. In real experiments, however, fluctuations of the differential energy shift $\delta$ coupling via the $\Jz$ term in the Hamiltonian~(\ref{eq.spin}) cause phase noise~\cite{Sinatra2012}.  
Both, the adiabatic and non-adiabatic approach to generate coherent spin squeezed states suffer from this noise problem. 
Especially the non-adiabatic method is very sensitive to DC phase noise (fluctuations from shot to shot). A spin echo technique~\cite{Vandersypen2004} can be used to effectively remove the sensitivity to these shot to shot fluctuations, however, the sensitivity to low but finite frequency fluctuations remains problematic. In general the spectral sensitivity is dependent on the duration of the free evolution and knowing the noise spectrum a specially designed spin echo technique can be used to optimize the spectral sensitivity. Figure~\ref{fig_phasenoise} shows an exemplary sensitivity curve versus frequency of the environmental noise with and without spin echo.  

\begin{figure}
	[t] 
	\begin{center}
		\includegraphics{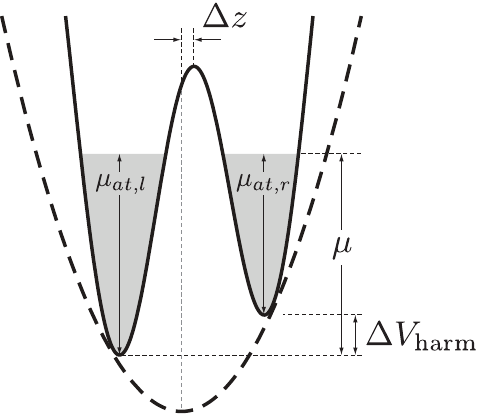} 
	\end{center}
	\caption{Position noise leads to relative atom number fluctuations. The schematic illustrates the connection of number difference fluctuations and relative position fluctuations $\Delta z$ between optical dipole trap and optical lattice. In equilibrium the chemical potential $\mu$ is the same in both wells which results in an occupation number difference between the left and right mode dependent on the differential energy shift $\Delta V_{\rm harm}$ between them.} \label{fig_trapShift} 
\end{figure}

In the adiabatic case the final relative atom number fluctuations are also influenced by the noise in the differential energy shift, however, the main contribution is due to shot to shot fluctuations.
The continuous coupling of the two states results effectively in continuous spin echo pulses counteracting dephasing.
The remaining DC  sensitivity can be understood intuitively for the example of the external double well system, but the argumentation also holds for an analogous internal system.
The system response to an imbalance in the offset energy for the two wells (a DC differential shift) is to adapt the relative atomic population such that the overall chemical potential is constant. This means an increased population of the lower state such that the extra interaction energy balances the external energy offset. An estimation for the noise sensitivity can be obtained by the following consideration where the main experimental challenge, the relative movement of the dipole traps that generate the trapping potential~\cite{Albiez2005a, Gati2006a} becomes obvious.
Figure~\ref{fig_trapShift} illustrates this situation. 

The optical dipole trap generates harmonic confinement in axial direction $V_{\rm harm} = m \omega_{z}^{2} z^{2}/2$, where $m$ is the atomic mass. Fluctuations of the energy difference $\Delta V_{\rm harm}$ between the two potential minima separated by $d$ due to fluctuations of the barrier position $\Delta z$ is given by $\Delta V_{\rm harm} = \pbp{V_{\rm harm}}{z} \Delta z = m \omega_{z}^{2} d \,\Delta z$.
The overall chemical potential in local density approximation~\cite{Pitaevskii2003} is $\mu = V_{\rm harm}(z) + \mu_{at}(z)$ and in equilibrium it is the same for the two wells. The contribution due to inter atomic interactions $\mu_{at}(z)$ balances the change in $V_{\rm harm}(z)$: $\Delta \mu_{at} = \pbp{\mu_{at}}{n_{a}} \Delta n_{a} = 2 \chi \Delta J_z$, where we used $\pbp{\mu_{at}}{n_{a}} = 2 \chi$~\cite{Zapata1998} and $J_z = (n_b-n_a)/2$.
To give an estimation of the required position stability of the trapping potentials we calculate the actual numbers for the trapping parameters of the double well trap in reference~\cite{Esteve2008}. The nonlinearity is $2 \chi \approx 2\pi\cdot1\,$Hz and the longitudinal trapping frequency $\omega_{z} \approx 2\pi \cdot 60\,$Hz. 
We obtain $\Delta z = \frac{2\chi}{m \omega_{z}^{2} d} \, \Delta J_z \lesssim 125 \mathrm{nm}$ for the position fluctuations leading to an extra noise of the same order as the shot noise limit $\xin = 0\,$dB for $2000$ atoms. Since these technical induced fluctuations add to the variance caused by the atomic quantum state, their magnitude has to be much smaller than the shot noise level in order to measure a reasonable amount of number squeezing. As a figure of merit, position fluctuations of $60\,$nm between different experimental realizations limit the best observable number squeezing to $\xin \approx -6\,$dB. This corresponds to roughly one percent of the double well spacing in this experiment~\cite{Esteve2008}.

\begin{figure*}
	[t] 
	\begin{center}
		\includegraphics{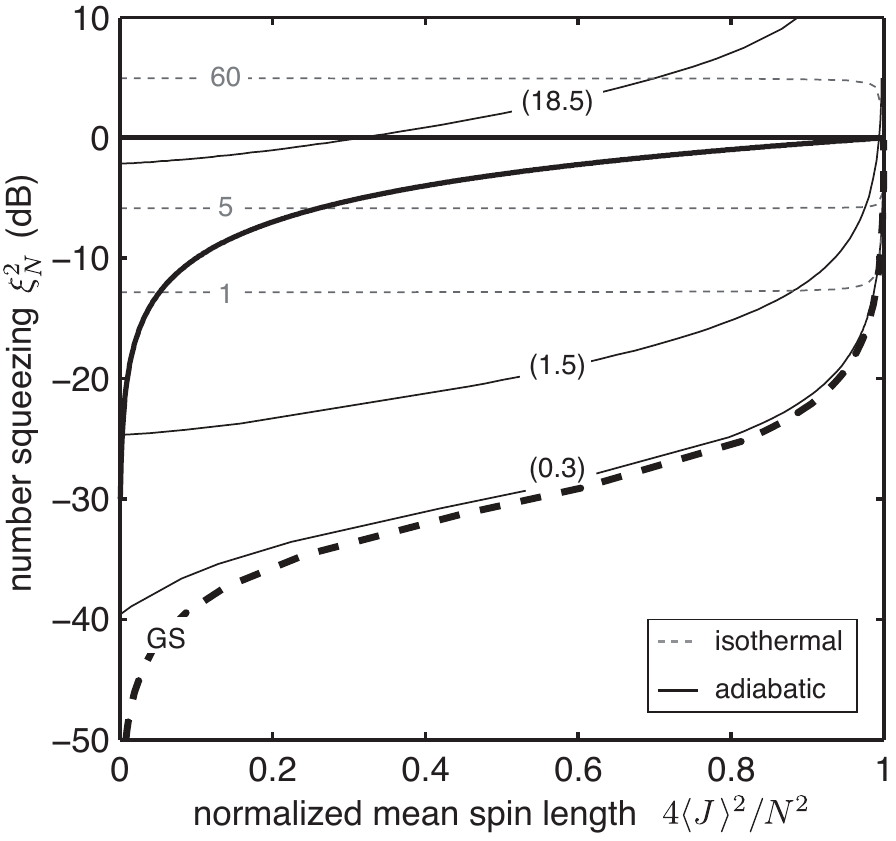} 
	\end{center}
	\caption{Adiabatic versus isothermal spin squeezing in a double well setup. Number squeezing versus mean spin length is shown for the same parameters of the experiment reported in~\cite{Esteve2008, Gross2010a, Gross2012} ($\chi = 2\pi\cdot 0.5\,$Hz, $N = 1600$, $\omega_z \approx 2\pi\cdot 65\,$Hz).   The solid thick black lines are the limits for number squeezing and coherent spin squeezing. The dashed thick black line corresponds to the ground state of the Hamiltonian~(\ref{eq.spin}). Light gray dashed lines are isothermals at different temperatures indicated on the left. The dark gray solid lines correspond to an adiabatic evolution of the system where the number of initially thermal populated many-body modes is given in brackets (corresponding to $60\,$nK, $5\,$nK and $1\,$nK initially). The regime parameter $\Lambda$ increases for each line from the right to the left.} \label{fig_adiabaticIsothermal} 
\end{figure*}

\subsection{Finite temperature.} 

The influence of finite temperature is very different in the two settings discussed here. While the effective temperature in the spin degree of freedom is nearly zero in the internal case, finite temperature effects easily dominate the fluctuations when dealing with a BEC in a double well trap~\cite{Pitaevskii2001, Gati2006}. 
The difference emerges from the very different initial energy gap between the two modes comprising the pseudo spin. When using two hyperfine states their energy separation is typically in the order of several gigahertz corresponding to temperatures of a few hundred millikelvin, much larger than the temperature in the BEC of a few tens of nanokelvin. Thus, using almost noise free coupling fields and short coupling pulses much faster than any thermalization timescale, spin states with almost zero temperature can be realized (see references~\cite{Sinatra2011, Sinatra2012} for the effect of finite temperature in this system.).

\begin{figure*}
	[t] 
	\begin{center}
		\includegraphics{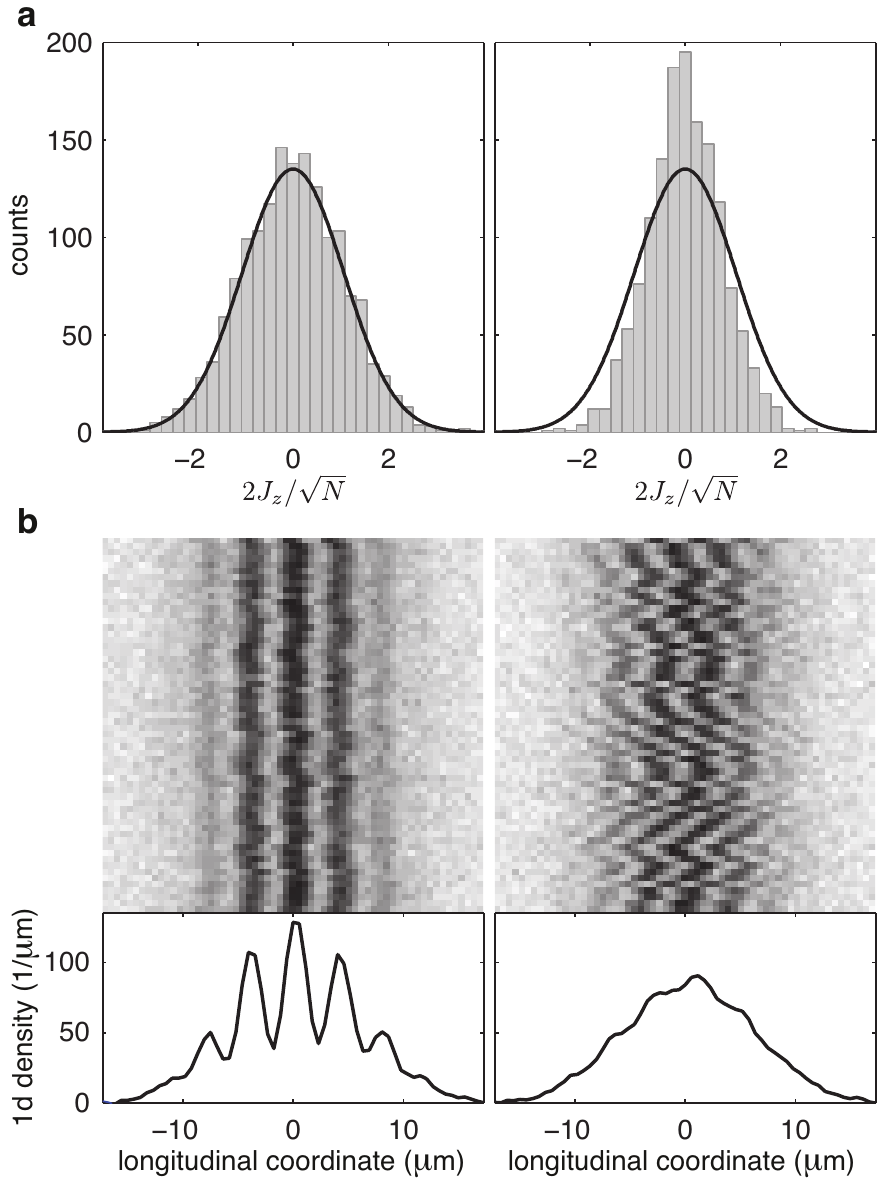} 
	\end{center}
	\caption{Number difference fluctuations and mean spin length. The figure shows the result of $1500$ measurements with experimental parameters chosen to minimize fluctuations in the relative phase (left column) or in the atom number difference (right column). {\bf a} Number difference fluctuations. The solid black line is a Gaussian fit to the left histogram. Its width is compatible with the shot noise noise limit taking detection noise into account. For comparison the solid black curve is copied to the right histogram whose width is approximately a factor of two smaller. {\bf b} The lower panel shows a vertical lineup of several measurements of the relative phase taken for the same experimental parameters as the number fluctuation measurements above. Each row is a horizontal profile of an independent measurement and the gray shading is proportional to the 1-dimensional atomic density. Below the ensemble average is shown revealing the mean spin length from the pattern visibility. Small variance in the atom number difference coincides with large variance in the relative phase (reduced mean spin length) and vice versa. This figure was originally published in~\cite{Gross2009}.} \label{fig_numberphase} 
\end{figure*}

The situation is very different in the external double well potential, where the pseudo spin in comprised of the two center of mass modes in the two potential wells. In a unbiased ($\delta = 0$) situation, there is no or a negligibly small energy gap separating the two modes such that several collective modes are excited even at nanokelvin temperatures. This leads to non pure initial spin states and to increased spin fluctuations as compared to a coherent spin state. 
Nevertheless the adiabatic spin squeezing technique results in an effective cooling of the spin fluctuations such that coherent spin squeezed states can be realized~\cite{Esteve2008} . Figure~\ref{fig_adiabaticIsothermal} shows the number fluctuations and mean spin lengths for different initial temperatures comparing the adiabatic method to the isothermal case. Parameters are the same as for the experiments reported in~\cite{Esteve2008}. Note that the fluctuations for the ground state would be orders of magnitude better for the same Hamiltonian parameters, showing the dominating effects of finite temperature.

\section{Examples of experimentally generated coherent spin squeezing.} 
In the previous sections we took two settings as examples for the generation of adiabatic and non-adiabatic spin squeezing in a BEC. We will briefly discuss some results of these experiments~\cite{Esteve2008, Gross2010} in the following. A more detailed description of these experiments can be found in~\cite{Gross2010a, Gross2012}. 

\begin{figure}
	[t] 
	\begin{center}
		\includegraphics{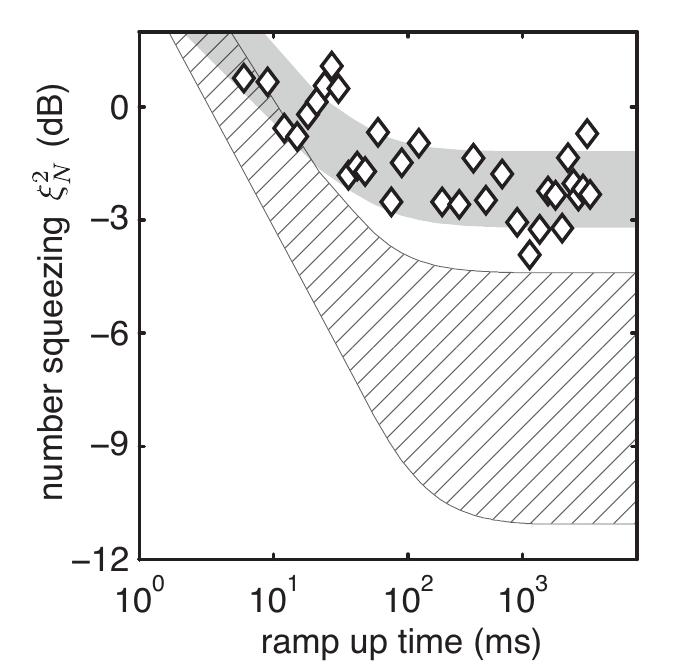} 
	\end{center}
	\caption{Timescales for adiabatic barrier ramps. To find the optimum ramp times the barrier was ramped linearly from a fixed start to a fixed end value with different speeds. The optimum is found where number squeezing starts to saturate. The hatched area, the theoretical two mode prediction for typical temperatures between $20\,$nK and $40\,$nK, cannot explain the observations. Including position noise (gray area) the measured data is reproduced by the theory for fluctuations of $\Delta z \approx 80\,$nm~\cite{Gross2010a, Gross2012}.} \label{fig_adiabaticity} 
\end{figure}

\subsection{Adiabatic spin squeezing in a double well potential.} The experiments discussed in this section have been originally published in~\cite{Esteve2008} and in~\cite{Gross2011}.
Approximately $2000$ condensed Rubidium atoms were loaded into a symmetric double trap where initially the potential barrier separating the two modes was small. For the initial parameters the system was already in the Josephson regime due to the rather strong nonlinearity. Next the potential barrier was increased adiabatically, driving the system deeper into the Josephson regime. The decrease of the tunnel coupling resulted in a decrease of the relative number fluctuations directly corresponding to the spin fluctuations in $J_z$ direction. The fluctuations were redistributed to the orthogonal spin direction such that the mean spin length decreased. The simultaneous change in the number fluctuations and mean spin length can be seen directly on the raw data (figure~\ref{fig_numberphase}).

Predicting the optimal time required for an adiabatic change of the parameters is not trivial in such an experiment, especially because additional aspects such as particle losses or heating limit the duration of the experiments. Therefore, the number squeezing was measured versus ramp up time of the double well potential while the final barrier height was kept constant. Figure~\ref{fig_adiabaticity} shows the results of this experiment. In order to explain the dependence of number squeezing versus ramp up time both finite temperature and finite fluctuations of the trap position had to be taken into account. They dominate the observed fluctuations (details of this experiment can be found in~\cite{Esteve2008, Gross2011, Gross2010a, Gross2012}). 

\begin{figure}[t] 
	\begin{center}
		\includegraphics{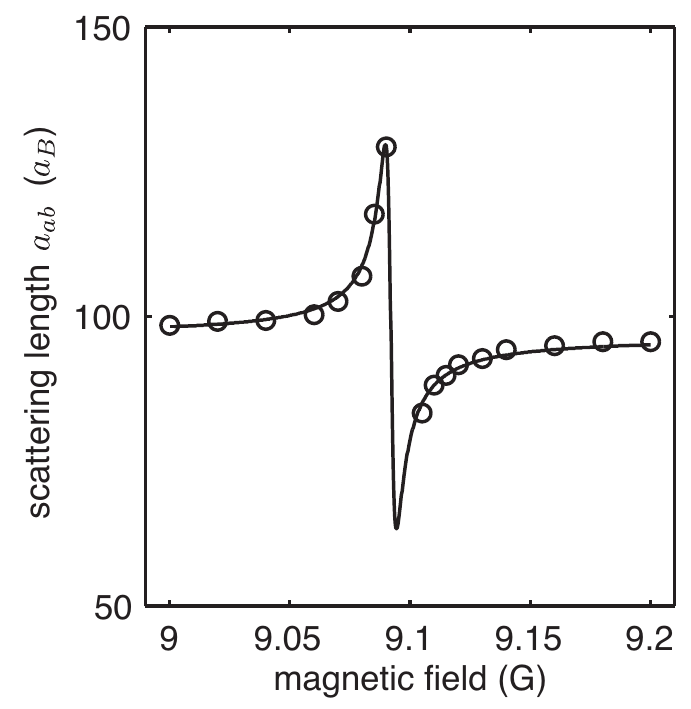} 
	\end{center}
	\caption{Shown is the measured interspecies scattering length $a_{ab}(B)$ around the Feshbach resonance~\cite{Gross2010a, Gross2012}. The black solid line is a fit to the real part of  $a_{ab}(B) = a_{ab}^{\mathrm{bg}}(1-\frac{\delta B}{(B-B_{0}-i \gamma_{B}/2})$~\cite{Chin2010} revealing  an inelastic width of $\gamma_{B}=4.6 \pm 0.7\,$mG in accordance with  the theoretical prediction~\cite{Kaufman2009}. The elastic width extracted from the fit is $\delta B = 1.6 \pm 0.2\,$mG. The center of the resonance is at $B_{0} = 9.092\,$G, but the absolute magnetic field has an uncertainty of approximately $10\,$mG in this measurement~\cite{Gross2010a, Gross2012}. The background scattering length $a_{ab}^{\mathrm{bg}}$ was chosen as a free parameter and the fit reveals $a_{ab}^{\mathrm{bg}}=96.5 \pm 0.7 \,a_{B}$.} \label{fig_feshbach} 
\end{figure}

Despite of these limitations coherent spin squeezed states in the double well trap were realized with a maximal squeezing factor of $\xis = -2.3\,$dB~\cite{Esteve2008}. However, the implementation of a full atom interferometer beyond the standard quantum limit remains an experimental challenge with such a setup even though the feasibility of the production of coherent spin squeezed states was demonstrated. The analog of electromagnetic coupling pulses rotating the spin state non-adiabatically needs to be implemented in order to realize a Ramsey sequence with the spin squeezed state as input state. The required non-adiabatic changes are difficult to realize without the possibility to change the nonlinearity directly (which could be done by using a Feshbach resonance~\cite{Chin2010}) since fast changes in the external trapping potential lead to unwanted excitation of the condensate. \\
Using internal hyperfine states to define the two modes instead of the center of mass modes these limitations can be overcome as presented in the following section.

\subsection{Non-adiabatic squeezing based on internal atomic states.} 

When working with internal hyperfine states of the atoms fast and precise spin rotations are readily available using electromagnetic coupling. The experimental challenge is to obtain a sufficiently high nonlinearity such that spin squeezed states can be realized within the experimental coherence time. Several ideas to increase the nonlinear interaction have been experimentally realized reaching from cavity mediated interactions in thermal gases~\cite{Schleier-Smith2010a, Leroux2010a} to state dependent microwave traps on atom chips which allow for the control of the spatial overlap of the two spin modes~\cite{Riedel2010}. 
We focus here on another method exploiting an interstate Feshbach resonance  between two hyperfine states of Rubidium around $9\,$G~\cite{Gross2010}~\footnote{The two states are the $F=1, m_F=1$ and $F=2, m_F=-1$ hyperfine states.}. 
Such a Feshbach resonance may be used to increase the nonlinearity but at the cost of increased particle losses~\cite{Chin2010}. 
Figure \ref{fig_feshbach} shows the measured resonantly enhanced effective nonlinearity~\cite{Zibold2010, Gross2010a, Gross2012}. 
\begin{figure*}[t] 
	\begin{center}
		\includegraphics{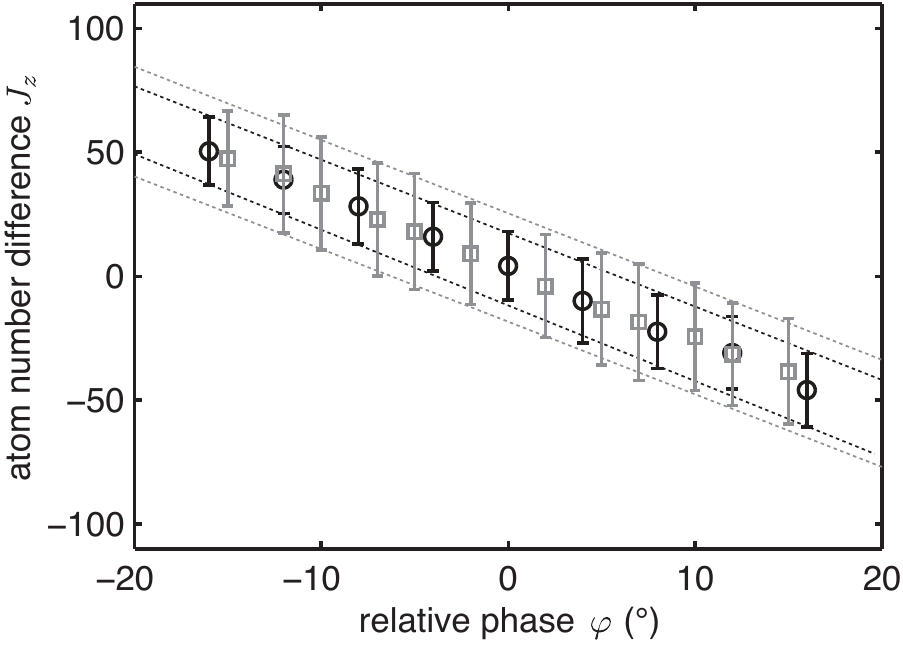} 
	\end{center}
	\caption{Linear versus nonlinear interferometric performance. Shown is the measured atom number difference versus relative phase after a Ramsey interferometric sequence. The error bars are two standard deviation statistical errors of the raw data. The black (gray) dashed lines are a fit through the upper and lower ends of the two standard deviation error bars for the non-linear (linear) interferometer. The horizontal width of the included areas measure the phase precision. The phase precision for the non-linear interferometer is increased by $31\%$ as compared to the linear one.} \label{fig_interferometer} 
\end{figure*}
The simultaneously enhanced losses limit the available time to generate coherently spin squeezed states to a few tens of milliseconds favoring the fast non-adiabatic ``one axis twisting'' method proposed in~\cite{Kitagawa1993} for the generation of spin squeezing~\cite{Li2008, Gross2010a, Gross2012}. Coherent spin squeezed states with inferred squeezing factors up to $\xis=-8.2\,$dB could be realized within an evolution time of $ 18\,$ms~\cite{Gross2010}. The use of a spin echo technique and an actively stabilized magnetic field~\cite{Zibold2010, Gross2010a, Gross2012} was crucial here in order to minimize the dephasing effects.

Given the possibility of precise spin rotations a full interferometric sequence can be implemented. In a prototypical measurement enhanced  interferometric sensitivity of $15\%$ beyond the standard quantum limit has been demonstrated~\cite{Gross2010} using coherent spin squeezed states within the Ramsey interferometer. 
Figure~\ref{fig_interferometer} shows a direct comparison of the Ramsey interferometer performance for coherent spin states as input state versus coherent spin squeezed states. In this direct comparison the performance increase due to the spin squeezing is even larger ($31\%$). However, it should be made clear that experimental noise is included here, that is, the comparison is not relative to the standard quantum limit. Measured is the increase in performance due to spin squeezing in this particular interferometric setup.

Several other experiments recently succeeded in the realization of quantum enhanced interferometers.
An interferometer with $1.1\,$dB precision gain has been realized using a cold sample of Cesium atoms where coherent spin squeezing has been generated by quantum non-demolition measurements~\cite{Louchet-Chauvet2010}.
Using cavity mediated interactions in a cold Rubidium gas to generate coherent spin squeezing the time dependent performance of a similar interferometer has been characterized in terms of the Allan variance~\cite{Leroux2010}. 

These different experiments show that atomic coherent spin squeezing can nowadays be realized with a variety of techniques and in quite different systems. It remains an open challenge to realize a large amount of coherent spin squeezing with atom numbers comparable to those used in modern linear interferometers. However, the basic concept of spin squeezing and its experimental feasibility has been demonstrated such spin squeezing might help to push the  limit of precision measurements in the future. 

\begin{acknowledgments}
I thank the BEC team at the Kirchhoff Institute for Physics in Heidelberg for their support and practical help. Especially I thank Markus Oberthaler for his great support and for his supervision of my PhD thesis. Special thanks goes to  Andreas Weller, Eike Nicklas, Tilman Zibold and J\'er\^ome Est\`eve for the joint work on our squeezing experiments. 
\end{acknowledgments}

\bibliographystyle{unsrt_mod}

\end{document}